\documentclass[a4paper]{llncs}
\usepackage{amssymb}
 \usepackage{amsfonts}
 \usepackage{amsmath}
\usepackage{float}
\usepackage{hyperref}
\usepackage{tikz}
\usetikzlibrary{arrows,shapes,decorations,automata,backgrounds,fit,matrix}
\usepackage{latexsym,calc}

\usepackage{stmaryrd}
\usepackage{xspace}
\usepackage{wrapfig}
\usepackage{multirow}
\usepackage{thm-restate}
\usepackage{paralist}
\usepackage{relsize}
\definecolor{aquamarine}{rgb}{0.50,1.00,0.83}%
\definecolor{violetred}{rgb}{0.82,0.13,0.56}%
\definecolor{Yellow}{rgb}{1,1,0}%
\definecolor{Peach}{rgb}{1,0.85,0.72}%
\definecolor{Blue}{rgb}{0,0,1}%
\definecolor{Red}{rgb}{1,0,0}%
\definecolor{White}{rgb}{1,1,1}%

\xdefinecolor{LemonChiffon}{rgb}{1,.98,.8}
\xdefinecolor{Vert}{rgb}{0.07,0.4,0.3}
\xdefinecolor{Green}{rgb}{0.0525,0.375,0.3}
\xdefinecolor{Bleu}{rgb}{0,.2,.54}
\xdefinecolor{Ver}{rgb}{0.0525,0.375,0.3}
\xdefinecolor{Rouge}{rgb}{1,0,0}
\xdefinecolor{Violet}{rgb}{.5,0,.7}
\xdefinecolor{Bordeaux}{rgb}{.7,.2,0}
\xdefinecolor{Ivory}{rgb}{1,1,0.9}
\xdefinecolor{Marron}{rgb}{0.6,0.4,0.4}
\xdefinecolor{GrisClair}{rgb}{0.75,0.75,0.75}
\xdefinecolor{GrisFonce}{rgb}{0.55,0.55,0.55}
\xdefinecolor{BleuMarine}{rgb}{0,0,0.5}
\xdefinecolor{Turquoise}{rgb}{0.4,1.,1.}
\xdefinecolor{Rose}{rgb}{1.0,0.8,0.8}
\xdefinecolor{RoseFonce}{rgb}{0.8,0.,0.6}
\xdefinecolor{BleuMarine}{rgb}{0,0,0.5}
\xdefinecolor{Blanc}{rgb}{0,0,0}
\xdefinecolor{Jaune}{rgb}{1,1,0}
\xdefinecolor{Noir}{rgb}{0,0,0}
\renewcommand{\iff}{\Leftrightarrow}

\DeclareMathOperator{\lcm}{lcm}

\newcommand{\abs}[1]{\mathrm{abs}(#1)}

\def\set#1{{\left\{ #1 \right\}}}
\def\tuple#1{{\langle #1 \rangle}}

\newcommand{\len}[1]{{|{#1}|}}

\newcommand{\card}[1]{\parallel\!\!{#1}\!\!\parallel}

\newcommand{\arrow}[2]{\xrightarrow[\scriptscriptstyle #2]{\scriptscriptstyle #1}}

\newcommand{\nat}{{\bf \mathbb{N}}}
\newcommand{\zed}{{\bf \mathbb{Z}}}

\newcommand{\rat}{{\bf \mathbb{Q}}}

\newcommand{\complex}{{\bf \mathbb{C}}}

\def\vr{\kern-\arraycolsep & \kern-\arraycolsep}
\def\VR{\kern-\arraycolsep\strut\vrule}

\newif\ifLongVersion\LongVersiontrue

\def\tuple#1{{\langle #1 \rangle}}

\newcommand{\idmatrix}{\mathbf{I}}

\newcommand{\Counters}{\mathsf{X}}
\newcommand{\acounter}{x}

\newcommand{\AtProps}{\mathsf{AP}}

\newcommand{\ConGuards}{\mathsf{CG}}
\newcommand{\aguard}{\mathsf{g}}

\newcommand{\States}{Q}
\newcommand{\astate}{q}
\newcommand{\Edges}{\Delta}
\newcommand{\anedge}{\delta}
\newcommand{\lab}{\Lambda}

\newcommand{\source}[1]{\mathit{source}(#1)}
\newcommand{\target}[1]{\mathit{target}(#1)}
\newcommand{\guard}[1]{\mathit{guard}(#1)}
\newcommand{\update}[1]{\mathit{update}(#1)}

\newcommand{\afunc}{f}

\newcommand{\AffineF}{\mathsf{Aff}}

\newcommand{\aconf}{\gamma}

\newcommand{\FlatMonoidAffineSystems}{\mathsf{AS}_{\mathrm{fm}}}
\newcommand{\TranslatingSystems}{\mathsf{TS}}
\newcommand{\FlatTranslatingSystems}{\mathsf{TS}_{\mathrm{f}}}
\newcommand{\FlatKripkeStructures}{\mathsf{KS}_{\mathrm{f}}}
\newcommand{\KripkeStructures}{\mathsf{KS}}

\newcommand{\arun}{\rho}

\newcommand{\amonoid}{\mathcal{M}}

\newcommand{\LTLnext}{{\mathsf X}}
\newcommand{\LTLuntil}{{\mathsf U}}
\newcommand{\LTLprevious}{{\mathsf X}^{-1}}
\newcommand{\LTLsince}{{\mathsf S}}
\newcommand{\LTLeventually}{{\mathsf F}}
\newcommand{\LTLalways}{{\mathsf G}}

\newcommand{\Vars}{{\mathsf{Var}}}
\newcommand{\avar}{{\mathsf z}}

\newcommand{\Reach}{\textsc{Reach}}

\newcommand{\MC}{\textsc{MC}}
\newcommand{\Logics}{\mathcal{L}}

\newcommand{\LTL}{\mathsf{LTL}}
\newcommand{\PLTL}{\mathsf{PLTL}}
\newcommand{\FO}{\mathsf{FO}}
\newcommand{\modelspltl}{\models_{\mathsf{\scriptscriptstyle PLTL}}}
\newcommand{\modelsfo}{\models_{\mathsf{\scriptscriptstyle FO}}}

\newcommand{\lengthof}[1]{\mathit{len}(#1)}
\newcommand{\sizeof}[1]{\mathit{size}(#1)}
\newcommand{\wordof}[1]{\mathit{trans}(#1)}
\newcommand{\ips}[1]{\mathit{ips}(#1)}
\newcommand{\polyone}[1]{\mathsf{Poly}_1(#1)}
\newcommand{\polytwo}[1]{\mathsf{Poly}_2(#1)}
\newcommand{\polythree}[1]{\mathsf{Poly}_3(#1)}
\newcommand{\poly}[1]{\mathsf{Poly}(#1)}

\newcommand{\anexec}{\theta}

\tikzstyle{init-left}=[pin={[pin distance=7pt,pin
  edge={<-,black,thick},pin position=left]}]

\begin{document}

\title{How hard is it to verify flat affine counter systems with the
  finite monoid property ?}

\author{Radu Iosif\inst{1} \and Arnaud Sangnier \inst{2}}
\institute{Verimag, Univ Grenoble Alpes, CNRS \and IRIF, Univ Paris Diderot, CNRS}

\maketitle
\begin{abstract}
We study several decision problems for counter systems with guards
defined by convex polyhedra and updates defined by affine
transformations. In general, the reachability problem is undecidable
for such systems. Decidability can be achieved by imposing two
restrictions: \begin{inparaenum}[(1)]
\item the control structure of the counter system is \emph{flat},
  meaning that nested loops are forbidden, and
\item the set of matrix powers is finite, for any affine update matrix
  in the system.
\end{inparaenum}
We provide tight complexity bounds for several decision problems of
such systems, by proving that reachability and model checking for Past
Linear Temporal Logic are complete for the second level of the
polynomial hierarchy $\Sigma^P_2$, while model checking for First
Order Logic is \textsc{PSPACE}-complete.
\end{abstract}

%%%%%%%%%%%%%%%%%%%%%%%%%%%%%%%%%%%%%%%%%%%%%%%%%%%%%%%%%%%%%%%%%%%%%%%%%%%%%%%%
\section{Introduction}
%%%%%%%%%%%%%%%%%%%%%%%%%%%%%%%%%%%%%%%%%%%%%%%%%%%%%%%%%%%%%%%%%%%%%%%%%%%%%%%%

Counter systems are finite state automata extended with integer
variables, also known as counter automata or counter machines. These
are Turing-complete models of computation, often used to describe the
behavior of complex real-life systems, such as embedded/control
hardware and/or software systems. Because many verification problems,
of rather complex systems, can be reduced to decision problems for
counter systems, it is important to understand the difficulties faced
by potential verification algorithms designed to work with the latter.

Due to their succintness and expressive power, most decision problems,
such as reachability, termination and temporal logic model-checking,
are undecidable for counter systems, even when the operations on the
counters are restricted to increment, decrement and zero-test
\cite{Minsky67}. This early negative result motivated the search for
subclasses with decidable decision problems. Such classes include
\emph{one-counter systems} \cite{goller-model-icalp10}, \emph{vector
  addition systems with states} \cite{Lipton76},
\emph{reversal-bounded counter machines} \cite{IbarraGurari81} and
\emph{flat counter systems} \cite{BoigelotPhD,FinkelLeroux02}.

Flat counter systems are defined by a natural syntactic restriction,
which requires that no state occurs in more than one simple cycle in
the control flow graph of the system. Decidability results on the
verification of reachability problems for flat counter systems have
been obtained by proving that, under certain restrictions on the logic
that defines the transition rules, the set of reachable configurations
is semilinear and effectively definable in Presburger arithmetic
\cite{BoigelotPhD,FinkelLeroux02,Cav10}. Even though flatness is an
important restriction (few counter system modeling real-life hardware
and software artifacts are actually flat), this class provides the
grounds for a useful method that under-approximates the set of
behaviors of a non-flat counter system by larger and larger sets of
paths described by flat counter systems. This method is currently used
by model checking tools, such as \textsc{Fast} \cite{Fast} and
\textsc{Flata} \cite{Flata}, and has been applied to improve the
results of static analysis \cite{MonniauxGawlitza12}, as well as the
convergence of counterexample-driven abstraction refinement algorithms
\cite{Atva12}. Moreover, several works define classes of
\emph{flattable} counter systems, for which there exist flat
unfoldings of the system with identical reachability sets. Such is the
case of timed automata\cite{ComonJurski99} and of 2-dimensional
vector addition systems with states\cite{LerouxSutre04,BlondinFinkelGoellerHaaseMcKenzie05}. For these
systems, the method of under-approximations by flat unfoldings is
guaranteed to terminate.

In general, the flatness restriction is shown to reduce the
computational complexity of several decision problems, such as
reachability or temporal logic model checking. For instance, in the
case of Kripke structures, flatness reduces the complexity of the
model-checking of Linear Temporal Logic ($\LTL$) from \textsc{PSPACE}
to \textsc{NP} \cite{kuhtz-weak-concur11}. When considering flat
counter systems whose updates are described by translations, the
complexity of these problems drops from undecidable to
\textsc{NP}-complete \cite{demri-taming-15}, while model checking for
First Order Logic ($\FO$) is coined to \textsc{PSPACE}-complete
\cite{demri-complexity-13}. For branching time temporal logics,
flatness yields decidable problems, but with less remarkable
complexity bounds \cite{demri-equivalence-rp14}.

In this work, we focus on the model of affine counter systems,
in which  each transition is labeled with \begin{inparaenum}[(i)]
\item a guard defined by convex polyhedra, i.e.\ linear systems of
  inequalities of the form $\vec{C} \cdot \vec{x} \leq \vec{d}$, and
\item deterministic updates defined by affine transformations given by
  a function $f(\vec{x})=\vec{A} \cdot \vec{x} + \vec{b}$
\end{inparaenum} where $\vec{A},\vec{C} \in \zed^{n \times n}$ are square matrices with
integer entries, $\vec{b},\vec{d} \in \zed^n$ are vectors of integer
constants and $\vec{x} = [x_1,\ldots,x_n]$ is a vector of
counters. For such systems, the set of reachable configurations is
semilinear (thus reachability is decidable), provided that the
multiplicative monoid generated by the matrices used in update
functions is finite. This condition is also known as the \emph{finite
  monoid property} \cite{BoigelotPhD,FinkelLeroux02}. Moreover, it
has been shown that the model-checking of such systems, for an
extended version of the branching time logic $\mathsf{CTL}^\ast$ is
decidable, also by reduction to the satisfiability of a Presburger
formula, of size exponential in the size of the counter system
\cite{demri-model-10}.

In this work, we show that for flat affine counter systems with the
finite monoid property, reachability and model checking for Past $\LTL$
are $\Sigma_2^P$-complete, whereas model checking for $\FO$ is
\textsc{PSPACE}-complete. Our result generalizes the results for flat
counter systems with translations
\cite{demri-complexity-13,demri-taming-15}, since these systems are a
strict subclass of flat affine counter
systems with the finite monoid property. For instance, a transfer of values
between different counters can be done in one step with an affine
counter system, whereas a translating counter system would need a
cycle to implement such operations. Our proof technique is based on an
analysis of the behavior of the sequence of matrix powers in a finite
multiplicative monoid, and adapts several techniques for translating
counter systems to this more general case.

\section{Counter Systems and their Decision Problems}
%%%%%%%%%%%%%%%%%%%%%%%%%%%%%%%%%%%%%%%%%%%%%%%%%%%%%%%%%%%%%%%%%%%%%%%%%%%%%%%%

We denote by $\nat$, $\zed$, $\rat$ and $\complex$ the sets of
natural, integer, rational and complex numbers, respectively. The
subscript $\geq0$, as in $\rat_{\geq0}$, indicates the set of positive
numbers, including zero. We write $[\ell,u]$ for the integer interval
$\set{\ell,\ell+1,\ldots,u}$, where $\ell\leq u$, $\abs{n}$ for the
absolute value of the integer $n\in\zed$ and $\lcm(n_1,\ldots,n_k)$
for the least common multiple of the natural numbers $n_1,\ldots,n_k
\in \nat$. The cardinality of a finite set $S$ is denoted by
$\card{S}$.

We denote by $\zed^{n \times m}$ the set of matrices with $n$ rows and
$m$ columns, where $\vec{A}[i]$ is the $i$-th column and
$\vec{A}[i][j]$ is the entry on the $i$-th row and $j$-th column of
$\vec{A} \in \zed^{n\times m}$, for each $i \in [1,n]$ and $j \in
[1,m]$. If $n=m$, we call this number the \emph{dimension} of $A$, and
we denote by $\idmatrix_n$ the identity matrix in $\zed^{n \times
  n}$. For $\vec{A} \in \zed^{n \times m}$ and $\vec{B} \in \zed^{m
  \times p}$, we denote by $\vec{A} \cdot \vec{B} \in \zed^{n \times
  p}$ the matrix product of $\vec{A}$ and $\vec{B}$. For a matrix
$\vec{A} \in \zed^{n \times n}$, we define $\vec{A}^0=\idmatrix_n$ and
$\vec{A}^i=\vec{A}^{i-1}\cdot \vec{A}$, for all $i>0$.

We write $\zed^n$ for $\zed^{n \times 1}$ in the following. Each
$\vec{v} \in \zed^n$ is a column vector, where $\vec{v}[i]$ is the
entry on its $i$-th row. For a vector $\vec{x}$ of variables of length
$n$ and a matrix $\vec{A} \in \zed^{m \times n}$, the product
$\vec{A}\cdot\vec{x}$ is the vector of terms $(\vec{A}\cdot
\vec{x})[i] = \sum_{j=1}^n \vec{A}[i][j]\cdot \vec{x}[j]$, for all
$i\in[1,m]$. A row vector is denoted by $\vec{v} = [v_1, \ldots, v_n]
\in \zed^{1 \times n}$. 
%We use $[\vec{A};\vec{B}]$ to denote the
%juxtaposition of two matrices $\vec{A} \in \zed^{m \times n}$ and
%$\vec{B} \in \zed^{m \times p}$. 
For a row vector $\vec{v}$, we denote
its transpose by $\vec{v}^\top$.

For a vector $\vec{v} \in \zed^n$, we consider the standard infinity
$\card{\vec{v}}_\infty = \max_{i=1}^n \abs{\vec{v}[i]}$ and euclidean
$\card{\vec{v}}_2 = \sqrt{\vec{v}\cdot\vec{v}^\top}$ norms. For a
matrix $\vec{A}\in\zed^{m \times n}$, we consider the induced norm
$\card{\vec{A}}_\infty = \max_{i=1}^m \sum_{j=1}^n
\abs{\vec{A}[i][j]}$. We also use the maximum matrix norm
$\card{\vec{A}}_{\max} = \max_{i=1}^m\max_{j=1}^n
\abs{\vec{A}[i][j]}$. The size of a matrix is
$\sizeof{\vec{A}}=\sum_{i=1}^m\sum_{j=1}^n \log_2(\vec{A}[i][j])$,
with integers encoded in binary.

\subsection{Counter systems}

Let $\Counters_n=\set{\acounter_1, \acounter_2, \ldots, \acounter_n}$
be a finite set of integer variables, called \emph{counters},
$\vec{x}$ be the vector such that $\vec{x}[i]=x_i$, for all $i \in
[1,n]$, and $\AtProps = \set{\mathrm{a},\mathrm{b},\mathrm{c},
  \ldots}$ be a countable set of boolean \emph{atomic propositions}. A
\emph{guard} is a (possibly empty) finite conjunction of inequalities
of the form $\vec{c} \cdot \vec{x} \leq d$, where $\vec{c} \in \zed^{1
  \times n}$, $d \in \zed$. We write $\top$ for the empty guard and
denote a non-empty guard by a system $\vec{C} \cdot \vec{x} \leq
\vec{d}$ of inequalities, where $\vec{C} \in \zed^{m \times n}$ and
$\vec{d} \in \zed^m$. An integer vector $\vec{v} \in \zed^n$
\emph{satisfies} the guard $\aguard$, written $\vec{v} \models
\aguard$, if either \begin{inparaenum}[(i)]
\item $\aguard \equiv \top$, or 
\item $\aguard \equiv \vec{C} \cdot \vec{x} \leq \vec{d}$ and
  $\vec{v}$ is a solution of the system.
\end{inparaenum}
The set of guards using $\Counters_n$ is denoted by
$\ConGuards(\Counters_n)$. An \emph{affine function} $\afunc : \zed^n
\rightarrow \zed^n$ is a pair $(\vec{A},\vec{b}) \in \zed^{n \times n}
\times \zed^n$. Given a vector $\vec{v} \in \zed^n$, the result of the
function $\afunc=(\vec{A},\vec{b})$ applied to $\vec{v}$ is
$\afunc(\vec{v})=\vec{A} \cdot \vec{v}+\vec{b}$. We
denote by $\AffineF_n$ the set of affine functions over $\zed^n$.

\begin{definition}\label{def:counter-systems}[Affine Counter System] 
For an integer $n \geq 0$, an \emph{affine counter system of dimension
  $n$} (shortly a counter system) is a tuple
$S=\tuple{\States,\Counters_n,\Edges,\lab}$,
where: \begin{inparaenum}[(i)]
\item $\States$ is a finite set of \emph{control states},
\item $\lab: \States \rightarrow 2^{\AtProps}$ is a \emph{labeling function}, and
\item $\Edges \subseteq \States \times \ConGuards(\Counters_n) \times
  \AffineF_n \times \States$ is a finite set of \emph{transition
    rules} labeled by guards and affine functions (updates).
\end{inparaenum}
\end{definition}

For a transition rule $\anedge=\tuple{\astate,\aguard,\afunc,\astate'}
\in \Edges$, we use the following notations $\source{\anedge}=\astate$,
$\guard{\anedge}=\aguard$, $\update{\anedge}=\afunc$ and
$\target{\anedge}=\astate'$. A \emph{path} $\pi$ of $S$
is a non-empty sequence of transition rules $\anedge_1 \ldots
\anedge_m$ such that $\source{\anedge_{i+1}}=\target{\anedge_i}$ for
all \(i \in [1,m-1]\). The path $\pi$ is a \emph{simple cycle} if
$\anedge_1 \ldots \anedge_m$ are pairwise distinct and
$\source{\anedge_1}=\target{\anedge_m}$. In this case, we denote
$\source{\pi}=\target{\pi}=\source{\anedge_1}$.  We say that a counter system $S$ is \emph{flat} if for each control state $q \in
\States$ there exists at most one simple cycle $\pi$ such that
$\source{\pi}=q$. Hence in  a flat counter system any path
leaving a simple cycle cannot revisit it.

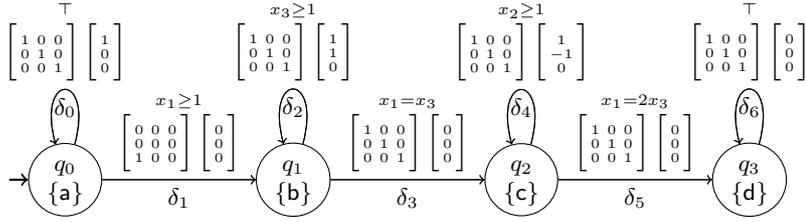
\begin{figure}[htbp]
\begin{center}

\begin{tikzpicture}[node distance=3cm]
\tikzstyle{every state}=[inner sep=0pt,minimum size=20pt]
\node(0)[state,init-left]{$\begin{array}{c}q_0 \\
                             \set{\mathsf{a}}\end{array}$}
edge [<-,loop above,distance=1cm] node[above] {$\begin{array}{c}
    \scriptstyle{\top} \\
     \left[\begin{array}{ccc}
          \scriptscriptstyle{1} & \scriptscriptstyle{0} & \scriptscriptstyle{0}\\[-2mm]
         \scriptscriptstyle{0} & \scriptscriptstyle{1} & \scriptscriptstyle{0}\\[-2mm]
         \scriptscriptstyle{0} & \scriptscriptstyle{0} & \scriptscriptstyle{1}
         \end{array}\right]
   \left[\begin{array}{c}
         \scriptscriptstyle{1} \\[-2mm]
      \scriptscriptstyle{0} \\[-2mm]
      \scriptscriptstyle{0}
     \end{array}\right]
   \end{array}$}
   (0)
edge [<-,loop above,distance=1cm] node[below] {$\delta_0$}(0);;

\node(1)[state,right of=0]{$\begin{array}{c}q_1 \\
                             \set{\mathsf{b}}\end{array}$} 
edge [<-] node[above] {$\begin{array}{c}
    \scriptstyle{ x_1\geq 1}  \\
     \left[\begin{array}{ccc}
          \scriptscriptstyle{0} & \scriptscriptstyle{0} & \scriptscriptstyle{0}\\[-2mm]
         \scriptscriptstyle{0} & \scriptscriptstyle{0} & \scriptscriptstyle{0}\\[-2mm]
         \scriptscriptstyle{1} & \scriptscriptstyle{0} & \scriptscriptstyle{0}
         \end{array}\right]
   \left[\begin{array}{c}
         \scriptscriptstyle{0} \\[-2mm]
      \scriptscriptstyle{0} \\[-2mm]
      \scriptscriptstyle{0}
     \end{array}\right]
   \end{array}$}
   (0)
edge [<-] node [below] {$\delta_1$}(0) 
edge [<-,loop above,distance=1cm] node[above] {$\begin{array}{c}
    \scriptstyle{x_3 \geq 1} \\
     \left[\begin{array}{ccc}
          \scriptscriptstyle{1} & \scriptscriptstyle{0} & \scriptscriptstyle{0}\\[-2mm]
         \scriptscriptstyle{0} & \scriptscriptstyle{1} & \scriptscriptstyle{0}\\[-2mm]
         \scriptscriptstyle{0} & \scriptscriptstyle{0} & \scriptscriptstyle{1}
         \end{array}\right]
   \left[\begin{array}{c}
         \scriptscriptstyle{1} \\[-2mm]
      \scriptscriptstyle{1} \\[-2mm]
      \scriptscriptstyle{0}
     \end{array}\right]
   \end{array}$}
   (1)
edge [<-,loop above,distance=1cm] node[below] {$\delta_2$}(1);
\node(2)[state,right of=1]{$\begin{array}{c}q_2 \\
                             \set{\mathsf{c}}\end{array}$} 
edge [<-] node[above] {$\begin{array}{c}
    \scriptstyle{x_1=x_3} \\
     \left[\begin{array}{ccc}
          \scriptscriptstyle{1} & \scriptscriptstyle{0} & \scriptscriptstyle{0}\\[-2mm]
         \scriptscriptstyle{0} & \scriptscriptstyle{1} & \scriptscriptstyle{0}\\[-2mm]
         \scriptscriptstyle{0} & \scriptscriptstyle{0} & \scriptscriptstyle{1}
         \end{array}\right]
   \left[\begin{array}{c}
         \scriptscriptstyle{0} \\[-2mm]
      \scriptscriptstyle{0} \\[-2mm]
      \scriptscriptstyle{0}
     \end{array}\right]
   \end{array}$}
   (1)
edge [<-] node [below] {$\delta_3$}(1) 
edge [<-,loop above,distance=1cm] node[above] {$\begin{array}{c}
    \scriptstyle{x_2 \geq 1} \\
     \left[\begin{array}{ccc}
          \scriptscriptstyle{1} & \scriptscriptstyle{0} & \scriptscriptstyle{0}\\[-2mm]
         \scriptscriptstyle{0} & \scriptscriptstyle{1} & \scriptscriptstyle{0}\\[-2mm]
         \scriptscriptstyle{0} & \scriptscriptstyle{0} & \scriptscriptstyle{1}
         \end{array}\right]
   \left[\begin{array}{c}
         \scriptscriptstyle{1} \\[-2mm]
      \scriptscriptstyle{-1} \\[-2mm]
      \scriptscriptstyle{0}
     \end{array}\right]
   \end{array}$}
   (2)
edge [<-,loop above,distance=1cm] node[below] {$\delta_4$}(2);
\node(3)[state,right of=2]{$\begin{array}{c}q_3 \\
                             \set{\mathsf{d}}\end{array}$} 
edge [<-] node[above] {$\begin{array}{c}
    \scriptstyle{x_1=2x_3} \\
     \left[\begin{array}{ccc}
          \scriptscriptstyle{1} & \scriptscriptstyle{0} & \scriptscriptstyle{0}\\[-2mm]
         \scriptscriptstyle{0} & \scriptscriptstyle{1} & \scriptscriptstyle{0}\\[-2mm]
         \scriptscriptstyle{0} & \scriptscriptstyle{0} & \scriptscriptstyle{1}
         \end{array}\right]
   \left[\begin{array}{c}
         \scriptscriptstyle{0} \\[-2mm]
      \scriptscriptstyle{0} \\[-2mm]
      \scriptscriptstyle{0}
     \end{array}\right]
   \end{array}$}
   (2)
edge [<-] node [below] {$\delta_5$}(2) 
edge [<-,loop above,distance=1cm] node[above] {$\begin{array}{c}
    \scriptstyle{\top} \\
     \left[\begin{array}{ccc}
          \scriptscriptstyle{1} & \scriptscriptstyle{0} & \scriptscriptstyle{0}\\[-2mm]
         \scriptscriptstyle{0} & \scriptscriptstyle{1} & \scriptscriptstyle{0}\\[-2mm]
         \scriptscriptstyle{0} & \scriptscriptstyle{0} & \scriptscriptstyle{1}
         \end{array}\right]
   \left[\begin{array}{c}
         \scriptscriptstyle{0} \\[-2mm]
      \scriptscriptstyle{0} \\[-2mm]
      \scriptscriptstyle{0}
     \end{array}\right]
   \end{array}$}
   (3)
edge [<-,loop above,distance=1cm] node[below] {$\delta_6$}(3);
\end{tikzpicture}
\end{center}
\caption{A flat affine counter system}
\label{fig:flat-cm}
\end{figure}

\begin{example}\label{ex:flat-counter-system}
Figure \ref{fig:flat-cm} shows a flat counter system whose control
states $\astate_0,\astate_1,\astate_2,\astate_3$ are labeled by the
atomic propositions $\mathsf{a},\mathsf{b},\mathsf{c},\mathsf{d}$,
respectively. From the initial state $\astate_0$ with all counters
equal to $0$, this system begins with incrementing $x_1$ a certain
number of times by a translation $\delta_0$ then, with $\delta_1$, it
transfers the value of the counter $x_1$ to $x_3$ and resets $x_1$;
the loop labeled by $\delta_2$ increments both $x_1$ and $x_2$ until
they both reach the value of $x_3$ and finally the loop labeled by
$\delta_4$ is used to decrement $x_2$ and increment $x_1$ until the
value of $x_1$ is twice the value of $x_3$. As a consequence, when the
system reaches $q_f$ the value of $x_1$ is twice the value of $x_3$
and the value of $x_2$ is equal to $0$.  Hence, any run reaching $\astate_3$
visits the state $\astate_1$ exactly the same number of times as the
state $\astate_2$. \hfill$\blacksquare$
\end{example}

The size of a counter system $S$ is
$\sizeof{S}=\sum_{\anedge\in\Edges} \sizeof{\anedge} +
\sum_{\astate\in\States}\card{\lab(\astate)}$, where $\sizeof{\anedge}
= 1 + \sizeof{\guard{\anedge}} + \sizeof{\update{\anedge}}$, for a
guard $\aguard\equiv\vec{C}\vec{x} \leq \vec{d}$ we have
$\sizeof{g}=\sizeof{\vec{C}}+\sizeof{\vec{d}}$, and for an update
$\afunc=(\vec{A},\vec{b})$, $\sizeof{f}=\sizeof{\vec{A}} +
\sizeof{\vec{b}}$.

A counter system of dimension $n=0$ is called a \emph{Kripke
  structure}. We denote by $\KripkeStructures$ and
$\FlatKripkeStructures$ the sets of Kripke structures and flat Kripke
structures, respectively. A counter system of dimension $n\geq1$ is
\emph{translating} if all updates labeling the transition rules are
pairs $(\idmatrix_n,\vec{b})$. Let $\TranslatingSystems$ and
$\FlatTranslatingSystems$ denote the sets of translating and flat
translating counter systems of any dimension $n\geq1$.

For a counter system $S$ of
dimension $n\geq1$, we consider
$\mathcal{M}_S\subseteq \zed^{n \times n}$ to be the smallest set of
matrices, closed under product, which contains $\idmatrix_n$ and each
matrix $\vec{A}$ occuring in an update $(\vec{A},\vec{b})$ of a
transition rule in $S$. Clearly, $\mathcal{M}_S$ forms a monoid with
the matrix product and identity $\idmatrix_n$. We say that $S$ has the
\emph{finite monoid property} if the set $\mathcal{M}_S$ is finite.
Let $\FlatMonoidAffineSystems$ be the set of flat counter
systems with the finite monoid property, which are the main focus of
this paper.

A \emph{configuration} of the counter system
$S=\tuple{\States,\Counters_n,\Edges,\lab}$ is a pair
$(\astate,\vec{v})\in\States\times\zed^n$, where $\astate$ is the
current control state and $\vec{v}[i]$ is the value of the counter
$x_i$, for all $i\in[1,n]$. Given two configurations
$\aconf=(\astate,\vec{v})$ and $\aconf'=(\astate',\vec{v}')$ and a
transition rule $\delta$, we write $\aconf \arrow{\anedge}{} \aconf'$
iff $\astate=\source{\anedge}$, $\astate'=\target{\anedge}$, $\vec{v}
\models \guard{\anedge}$ and $\vec{v}'=\update{\anedge}(\vec{v})$. We
use the notation $\aconf \arrow{}{} \aconf'$ when there exists a
transition rule $\anedge$ such that $\aconf \arrow{\anedge}{}
\aconf'$. A \emph{run} of $S$ is then an infinite sequence of the form
$\arun ~\colon~ \aconf_0 \arrow{\delta_0}{} \aconf_1
\arrow{\delta_1}{} \aconf_2 \arrow{\delta_2}{} \ldots$. We say that
such a run starts at configuration $\aconf_0$, furthermore we denote
by $\wordof{\arun}=\delta_0\delta_1\delta_2\ldots$ the infinite
sequence of transition rules seen during $\arun$. Without loss of
generality we consider \emph{deadlock-free} counter systems only,
where for each configuration $\aconf \in \States\times\zed^n$, there
exists a configuration $\aconf'$ such that $\aconf \arrow{}{}
\aconf'$\footnote{We ensure deadlock-freedom by adding a sink state
  $\sigma$ to $S$, with a self-loop $\sigma \arrow{\top}{} \sigma$,
  and a transition $\astate \arrow{\top}{} \sigma$ from each state
  $\astate \in \States$.}.

\begin{example}\label{ex:run}
The sequence below is a run of the counter system from Figure
\ref{fig:flat-cm}: 
\[\begin{array}{l}
\left(q_0,\left[\begin{array}{c}0\\0\\0\end{array}\right]\right) \arrow{\anedge_0}{}
\left(q_0,\left[\begin{array}{c}1\\0\\0\end{array}\right]\right) \arrow{\anedge_1}{}
\left(q_1,\left[\begin{array}{c}0\\0\\1\end{array}\right]\right) \arrow{\anedge_2}{}
\left(q_1,\left[\begin{array}{c}1\\1\\1\end{array}\right]\right) \arrow{\anedge_3}{}
\left(q_2,\left[\begin{array}{c}1\\1\\1\end{array}\right]\right) \\\\
\arrow{\anedge_4}{} \left(q_2,\left[\begin{array}{c}2\\0\\1\end{array}\right]\right) 
\arrow{\anedge_5}{} \left(q_3,\left[\begin{array}{c}2\\0\\1\end{array}\right]\right) 
\arrow{\anedge_6}{}
    \left(q_3,\left[\begin{array}{c}2\\0\\1\end{array}\right]\right) 
\arrow{\anedge_6}{} \left(q_3,\left[\begin{array}{c}2\\0\\1\end{array}\right]\right) \arrow{\anedge_6}{}\cdots
\hfill\blacksquare
\end{array}\]
\end{example}

\subsection{Decision Problems}

The \emph{reachability problem} for a class of counter systems
$\mathcal{C}$, denoted by $\Reach(\mathcal{C})$, can then be stated as
follows: given a counter system $S$ in $\mathcal{C}$, an initial
configuration $\aconf_0$, and a control state $\astate_f$, does $S$
have a run starting in $\aconf_0$ and containing a configuration
$(\astate_f,\vec{v})$, for some $\vec{v} \in \zed^n$? It is well known
that $\Reach(\TranslatingSystems)$ is undecidable for non-flat counter
systems, even for only $2$ counters with zero test guards, and
increment/decrement updates \cite{Minsky67}.

In this work we also consider \emph{model checking} problems for two
specification logics, namely Past Linear Temporal Logic ($\PLTL$) and
First Order Logic ($\FO$). The formulae of $\PLTL$ are defined by the
grammar: $\phi ::= \mathsf{p} \mid \neg \phi \mid \phi \wedge \phi
\mid \LTLnext \phi \mid \phi \LTLuntil \phi \mid \LTLprevious \phi
\mid \phi\LTLsince \phi$, where $\mathsf{p} \in \AtProps$. As usual,
we consider the derived modal operators $\LTLeventually \phi \equiv
\phi \LTLuntil \top$ and $\LTLalways \phi \equiv
\neg\LTLeventually\neg\phi$. Given a run $\arun \colon~ \aconf_0
\arrow{\delta_0}{} \aconf_1 \arrow{\delta_1}{} \aconf_2
\arrow{\delta_2}{} \ldots$ of a counter system $S$ and a $\PLTL$
formula $\phi$, the semantics of $\PLTL$ is defined by an inductive
forcing relation $\rho,i \modelspltl \phi$, where for all $i\geq0$:
\[\begin{array}{rcl}
\rho,i \modelspltl \mathsf{p} & \iff & \gamma_i=(\astate,\vec{v}) \text{ and } \mathsf{p} \in \lab(\astate) \\
\rho,i \modelspltl \LTLnext\phi & \iff & \rho,i+1 \modelspltl \phi \\
\rho,i \modelspltl \phi\LTLuntil\psi & \iff & \rho,j \modelspltl \psi 
\text{ for some } j\geq i \text{ and } \rho,k \modelspltl \phi \text{ for all } i \leq k < j \\
\rho,i \modelspltl \LTLprevious\phi & \iff & i>0 \text{ and } \rho,i-1 \modelspltl \phi \\
\rho,i \modelspltl \phi\LTLsince\psi & \iff & \rho,j \modelspltl \psi \text{ for some } 0 \leq j \leq i 
\text{ and } \rho,k \modelspltl \phi \text{ for all } j < k \leq  i \\
\end{array}\]
The semantics of the boolean connectives $\wedge$ and $\neg$ is the
usual one. We write $\rho \modelspltl \phi$ for $\rho,0 \modelspltl
\phi$. For instance, each run of the counter system from Figure
\ref{fig:flat-cm} satisfies the $\PLTL$ formula $\LTLalways((\mathsf{b}
\wedge \LTLnext \mathsf{b} \wedge \LTLeventually \mathsf{d})\rightarrow \LTLeventually (\mathsf{c}
\wedge \LTLnext \mathsf{c}))$, because each run visiting $\astate_3$ sees the same number
of $\mathsf{b}$'s and $\mathsf{c}$'s. 

The formulae of $\FO$ are defined by the grammar: $\phi ::=
\mathsf{p}(\avar) \mid \avar < \avar' \mid\neg \phi \mid \phi \wedge
\phi \mid\exists\avar. \phi$, where $\mathsf{p} \in \AtProps$ and
$\avar$ belongs to a countable set of \emph{logical variables}
$\Vars$. The semantics is given by a forcing relation $\rho \modelsfo
\phi$ between runs $\rho$ of $S$ and closed formulae $\phi$, with no
free variables, which interprets the quantified variables
$\avar\in\Vars$ as positive integers denoting positions in the
run. With this convention, the semantics of $\FO$ is standard. For
instance, each run of the counter system from Figure \ref{fig:flat-cm}
satisfies the $\FO$ property: $\forall\mathsf{x}\forall\mathsf{x}' .
(\mathsf{x} < \mathsf{x}' \wedge \mathsf{b}(\mathsf{x})  \wedge
\mathsf{b}(\mathsf{x}') \wedge \exists\mathsf{z}.\mathsf{d}(\mathsf{z}))\rightarrow
\exists\mathsf{y}\exists\mathsf{y}' ~.~ \mathsf{c}(\mathsf{y}) \wedge
\mathsf{c}(\mathsf{y}')$, which differs from the previous $\PLTL$
formula only in that $\mathsf{x}$ and $\mathsf{x}'$ ($\mathsf{y}$ and
$\mathsf{y}'$) are not necessarily consecutive moments in time.

The model-checking problem for counter systems in a class
$\mathcal{C}$ with specification language $\Logics$ (in this work
either $\PLTL$ or $\FO$), denoted by $\MC_\Logics(\mathcal{C})$, is
defined as follows: given a counter system $S$ in $\mathcal{C}$, an
initial configuration $\aconf_0$, and a formula $\phi$ of $\Logics$,
does there exist a run $\rho$ of $S$ starting in $\aconf_0$ such that
$\rho \models_\Logics \phi$.

\begin{table}[htb]
\begin{center}
\scalebox{0.9}{
\begin{tabular}{|c|c|c|c|c|c|}
\hline
&~$\KripkeStructures$~&~$\FlatKripkeStructures$~&~ $\TranslatingSystems$~ &~
$\FlatTranslatingSystems$ ~&~$\FlatMonoidAffineSystems$~ \\
\hline\hline
$\Reach$ &~\textsc{NLOGSPACE}~&~\textsc{NLOGSPACE}~&~Undec. \cite{Minsky67} &~
\textsc{NP}-c.\cite{demri-taming-15} ~&~\textsc{4EXPtime} \cite{FinkelLeroux02}~\\
\hline
$\MC_\PLTL$ &~\textsc{PSPACE}-c.\cite{sistla-complexity-85}~&~\textsc{NP}-c.\cite{demri-taming-15,kuhtz-weak-concur11}~&~Undec. 
&~\textsc{NP}-c. \cite{demri-taming-15} ~&~\textsc{4EXPTIME} \cite{demri-model-10}~\\
\hline
$\MC_\FO$ &~\textsc{NONELEM.} \cite{stockmeyer-complexity-74}~&~\textsc{PSPACE}-c. \cite{demri-complexity-13}~&~Undec. 
&~\textsc{PSPACE}-c. \cite{demri-complexity-13} ~&~Decid. \cite{demri-model-10}~\\
\hline
\end{tabular}
}
\end{center}
\caption{Known results}
\label{tab:known-results}
\end{table}

Table \ref{tab:known-results} 
gives an overview of the known
complexity bounds for the previously mentioned decision problems
looking at different classes of counter systems\footnote{The
  \textsc{NP}-hardness for the
  reachability problem in flat translating counter systems was
  obtained in \cite{demri-taming-15} for
  systems equipped  with more
  expressive guards, but the reduction  can be adapted to the case of
  conjunctive guards.}. 
For flat Kripke structures, it is proved in \cite{demri-taming-15,kuhtz-weak-concur11}
that $\MC_\PLTL(\FlatKripkeStructures)$ is \textsc{NP}-complete and in
\cite{demri-complexity-13} that $\MC_\FO(\FlatKripkeStructures)$ is
\textsc{Pspace}-complete, whereas $\MC_\PLTL(\KripkeStructures)$ is
\textsc{Pspace}-complete and $\MC_\FO(\KripkeStructures)$ is
non-elementary. As explained in
\cite{demri-complexity-13,demri-taming-15}, the complexity of these
two last problems does not change if one considers flat translating
counter systems. For what concerns flat counter systems with the
finite monoid property, it has been shown that one can compute a
Presburger formula which characterizes the reachability set, which
entails the decidability of $\Reach(\FlatMonoidAffineSystems)$
\cite{FinkelLeroux02}. Later on, in \cite{demri-model-10}, the authors have
shown that the model-checking of an extension of the branching time
logic $\mathsf{CTL}^\ast$ is decidable. Hence we know that
$\MC_\PLTL(\FlatMonoidAffineSystems)$ and
$\MC_\FO(\FlatMonoidAffineSystems)$ are decidable, however no precise
complexity for these problems is known. We only can deduce from the
proofs in \cite{FinkelLeroux02,BIKLmcs14,demri-model-10} that for
$\Reach(\FlatMonoidAffineSystems)$
and $\MC_\PLTL(\FlatMonoidAffineSystems)$ there exists a reduction to
the satisfiability problem for Presburger arithmetic where the built
formula is exponentially bigger than the size of the model, this leads
to an upper bound in \textsc{4EXPTIME} (the satisfiability problem for
Presburger arithmetic can in fact be solved in \textsc{3EXPTIME}, see
e.g.\ \cite{haase-subclasses-14}).

In this work, we aim at providing the exact complexity for the
problems related to affine counter systems with the finite monoid
property, in other words to improve the last column of the
table. Finally, note that for the presented results, the counter
systems were manipulating natural numbers instead of integers, but
considering the latter option does not change the stated results.

%%%%%%%%%%%%%%%%%%%%%%%%%%%%%%%%%%%%%%%%%%%%%%%%%%%%%%%%%%%%%%%%%%%%%%%%%%%%%%%%
\section{A Hardness Result}
\label{sec:lower-bounds}
%%%%%%%%%%%%%%%%%%%%%%%%%%%%%%%%%%%%%%%%%%%%%%%%%%%%%%%%%%%%%%%%%%%%%%%%%%%%%%%%

In this section we prove that the reachability problem for flat affine
counter systems with the finite monoid property is $\Sigma_2^P$-hard,
by reduction from the validity problem for the $\exists^*\forall^*$
fragment of quantified boolean formulae ($\Sigma_2$-QBF), which is a
well-known $\Sigma_2^P$-complete problem \cite[\S
  5.2]{AroraBarakBook}. Let us consider a formula $\Phi \equiv \exists
y_1 \ldots \exists y_p \forall z_1 \ldots \forall z_q ~.~
\Psi(\vec{y}, \vec{z})$, where $\vec{y} = \set{y_1, \ldots, y_p}$ and
$\vec{z} = \set{z_1, \ldots, z_q}$ are non-empty sets of boolean
variables, and $\Psi$ is a quantifier-free boolean formula in
conjunctive normal form (CNF) --- this choice does not affect the
hardness of the problem. We shall build, in polynomial time, a flat
counter system $S_\Phi$, with the finite monoid property, such that
$\Phi$ is valid if and only if $S_\Phi$ has a run reaching $\astate_f$
which starts in $(\astate_0,\vec{v}_0)$ for a
certain valuation $\vec{v}_0$ of its counters.

\begin{figure}[htbp]
\vspace*{-\baselineskip}
\begin{center}
\scalebox{0.8}{
\begin{tikzpicture}[node distance=1.7cm]
\tikzstyle{every state}=[inner sep=0pt,minimum size=20pt]
\node(0)[state,init-left]{$q_0$};
\node(1)[state,right of=0]{$q_1$} 
  edge [<-,thick,out=135,in=45] node[above] {$(\vec{I}_N,\vec{e}_1)$}
  (0)
  edge [<-,thick,out=-135,in=-45] node[below] {$(\vec{I}_N,\vec{0})$}
  (0);
\node(2)[state,right of=1]{$q_2$} 
  edge [<-,thick,out=135,in=45] node[above] {$(\vec{I}_N,\vec{e}_2)$}
  (1)
  edge [<-,thick,out=-135,in=-45] node[below] {$(\vec{I}_N,\vec{0})$}
  (1);
\node(3)[state,right of=2]{$q_3$} 
  edge [<-,thick,out=135,in=45] node[above] {$(\vec{I}_N,\vec{e}_3)$}
  (2)
  edge [<-,thick,out=-135,in=-45] node[below] {$(\vec{I}_N,\vec{0})$}
  (2);
\node(tmp)[right of=3,xshift=-3em]{$\cdots$};
\node(p-1)[state,right of=tmp,xshift=-3em]{$q_{p-1}$};
\node(p)[state,right of=p-1]{$q_p$} 
  edge [<-,thick,out=135,in=45] node[above] {$(\vec{I}_N,\vec{e}_p)$}
  (p-1)
  edge [<-,thick,out=-135,in=-45] node[below] {$(\vec{I}_N,\vec{0})$}
  (p-1);
\node(q)[state,right of=p,xshift=1em]{$q$} 
  edge [<-,thick] node[above] {$\begin{array}{c} g_1 \\(\vec{M},\vec{0})\end{array}$}
  (p)
  edge [<-,thick,loop above] node[above] {$\begin{array}{c} g_1 \\(\vec{M},\vec{0})\end{array}$}
  (q);
\node(qf)[state,right of=q,xshift=1em]{$q_f$} 
  edge [<-,thick] node[above] {$\begin{array}{c} g_2 \\ (\vec{I}_N,\vec{0})\end{array}$}(q);

\end{tikzpicture}
}
\end{center}
\vspace*{-\baselineskip}
\caption{The counter system $S_\Phi$ corresponding to the
  $\Sigma_2$-QBF $\Phi$}
\label{fig:reduc-first-phase}
\end{figure}
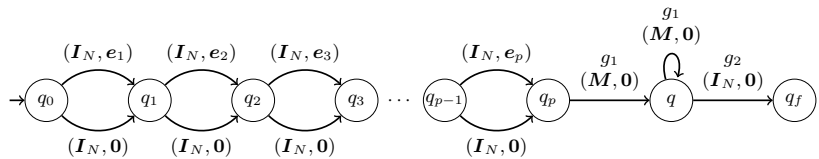

Let $\pi_n$ denote the $n$-th prime number, i.e.\ $\pi_1=2, \pi_2=3,
\pi_3=5$, etc. Formally, $S_\Phi =
\tuple{\States,\Counters_N,\Edges,\lab}$, where
$\States=\set{q_0,\ldots,q_p,q,q_f}$, $N = p + \sum_{n=1}^q \pi_n$,
and $\lab$ is the function associating to  each state an empty set of
propositions. The transition rules $\Edges$ are depicted in Figure
\ref{fig:reduc-first-phase}. Intuitively, each existentially
quantified boolean variable $y_i$ of $\Phi$ is modeled by the counter
$x_i$ in $S_\Phi$, each universally quantified variable $z_j$ of
$\Phi$ is modeled by the counter $x_{p+\sum_{n=1}^{j}\pi_n}$, and the
rest are working counters. All counters range over the set
$\set{0,1}$, with the obvious meaning ($0$ stands for false and $1$
for true).

The counter system $S_\Phi$ works in two phases. The first phase,
corresponding to transitions $q_0 \arrow{}{} \ldots \arrow{}{} q_p$,
initializes the counters $x_1, \ldots, x_p$ to some values from the
set $\set{0,1}$, thus mimicking a choice of boolean values for the
existentially quantified variables $y_1, \ldots, y_p$ from
$\Phi$. Here $\idmatrix_N \in \zed^{N \times N}$ is the identity
matrix, and $\vec{e}_i \in \set{0,1}^N$ is the unit vector such that
$\vec{e}_i[j]=0$ if $j\neq i$ and $\vec{e}_i[i]=1$.

\begin{figure}[htbp]
\scriptsize

\begin{center}

$
\begin{array}{ccccc}
\begin{pmatrix}

\begin{tikzpicture}[every node/.style={minimum width=1em}]

\node(a)[scale=1.5] {$\vec{I}_{p}$};
\matrix (m1)  [matrix of math nodes]
{ 
 \textcolor{white}{0} &  \textcolor{white}{0} & \textcolor{white}{0} \\
 \textcolor{white}{0} &  \textcolor{white}{0} & \textcolor{white}{0} \\
 \textcolor{white}{0} &  \textcolor{white}{0}& \textcolor{white}{0} \\
};
\matrix (m2) at (m1-3-3.south east) [anchor=m2-1-1.north west,
matrix of math nodes]
{ 
 \textcolor{white}{0} &  \textcolor{white}{0} & \textcolor{white}{0} \\
 \textcolor{white}{0} &  \textcolor{white}{0} & \textcolor{white}{0} \\
 \textcolor{white}{0} &  \textcolor{white}{0}& \textcolor{white}{0} \\
};
\node(b)[scale=1.5,xshift=1.2em,yshift=-1.5em] at (m1-3-3.south east) {$\vec{M}_{\pi_1}$};

\matrix (m3) at (m2-3-3.south east) [anchor=m3-1-1.north west,
matrix of math nodes]
{ 
 \textcolor{white}{0} &  \textcolor{white}{0} \\

 \textcolor{white}{0} &  \textcolor{white}{0} \\
};
\node(b)[scale=1.5,xshift=1.3em,yshift=-0.7em] at (m2-3-3.south east) {$\ddots$};

\matrix (m4) at (m3-2-2.south east) [anchor=m4-1-1.north west,
matrix of math nodes]
{ 
 \textcolor{white}{0} &  \textcolor{white}{0} & \textcolor{white}{0} \\
 \textcolor{white}{0} &  \textcolor{white}{0} & \textcolor{white}{0} \\
 \textcolor{white}{0} &  \textcolor{white}{0}& \textcolor{white}{0} \\
};
\node(d)[scale=1.5,xshift=1.5em,yshift=-1.5em] at (m3-2-2.south east)
{$\vec{M}_{\pi_q}$};

\node[scale=4,xshift=0.2em] at (m1 |- m3) {$0$};
\node[scale=4,yshift=0.2em] at (m4 |- m2) {$0$};

\node(1) [xshift=5pt, yshift=2pt] at (m1-1-3.north east) {};
\node(2)  [xshift=5pt,yshift=-2pt] at (m1-3-3.south east) {}
edge [<->] node [right] {$p$} (1);
\node(3) [xshift=-2pt, yshift=-5pt] at (m1-3-1.south west) {};
\node(4)  [xshift=2pt,yshift=-5pt] at (m1-3-3.south east) {}
edge [<->] node [below] {$p$} (3);

\node(5) [xshift=5pt, yshift=2pt] at (m2-1-3.north east) {};
\node(6)  [xshift=5pt,yshift=-2pt] at (m2-3-3.south east) {}
edge [<->] node [right] {$\pi_1$} (5);
\node(7) [xshift=-2pt, yshift=-5pt] at (m2-3-1.south west) {};
\node(8)  [xshift=2pt,yshift=-5pt] at (m2-3-3.south east) {}
edge [<->] node [below] {$\pi_1$} (7);

\node(9) [xshift=-5pt, yshift=2pt] at (m4-1-1.north west) {};
\node(10)  [xshift=-5pt,yshift=-2pt] at (m4-3-1.south west) {}
edge [<->] node [left] {$\pi_q$} (9);
\node(11) [xshift=-2pt, yshift=5pt] at (m4-1-1.north west) {};
\node(12)  [xshift=2pt,yshift=5pt] at (m4-1-3.north east) {}
edge [<->] node [above] {$\pi_q$} (11);

\draw (m1-1-3.north east) -- (m1-1-3.north east |- m2-3-1.south west);
\draw (m1-3-1.south west) -- (m2-1-3.north east);
\draw (m2-1-3.north east) -- (m2-3-3.south east);
\draw (m2-3-1.south west) -- (m2-3-3.south east);
\draw (m4-1-1.north west) -- (m4-1-3.north east);
\draw (m4-1-1.north west) -- (m4-3-1.south west);
\end{tikzpicture}
\end{pmatrix} 
& \hspace*{1.5cm} & 
\begin{pmatrix}
\begin{tikzpicture}[every node/.style={minimum width=1em}]
\matrix (m1)  [matrix of math nodes]
{ 
 0 &  1 & \textcolor{white}{0}   &  \textcolor{white}{0}   &\textcolor{white}{0} \\
 \textcolor{white}{0} & 0  & 1 & \ddots & \textcolor{white}{0}  \\
 \textcolor{white}{0}  &  \textcolor{white}{0}  & \ddots &
 \ddots & \textcolor{white}{0}  \\
0 &  \textcolor{white}{0} & \textcolor{white}{0}  &  0 & 1 \\
 1 &  \textcolor{white}{0} & \cdots &  \textcolor{white}{0} & 0 \\
};
\node[scale=4,xshift=0.8em,yshift=-0.2em] at (m1-1-2.south east) {$0$};
\end{tikzpicture}
\end{pmatrix}
& \hspace*{1.5cm} &
\begin{pmatrix}
\begin{tikzpicture}[every node/.style={minimum width=1}]
\matrix (m1)  [matrix of math nodes]
{ 
 0\\ 
\vdots \\
0\\
\hline
0\\
1\\
\hline
\vdots \\
\hline
0\\ 
\vdots \\
0\\
1\\
};
\end{tikzpicture}
\end{pmatrix}
\\
\vec{M} && \vec{M}_{\pi_j} && \vec{v}_0
\end{array}
$
\end{center}

\caption{Matrix $\vec{M}$ and initial vector $\vec{v}_0$}
\label{fig:mat-reduction}
\end{figure}
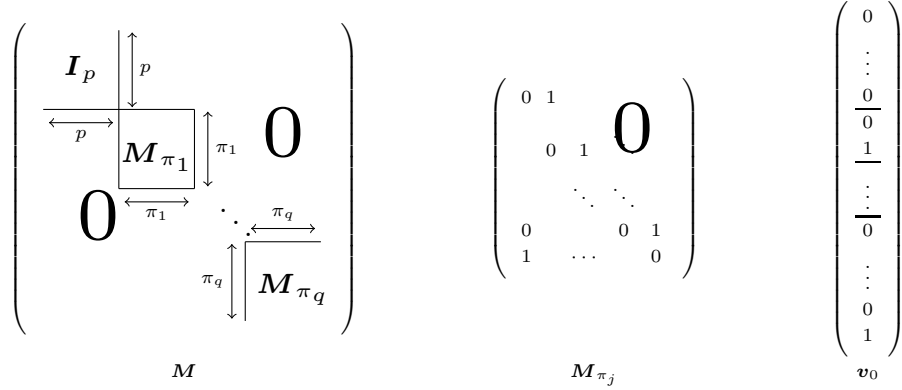

The second phase checks that $\Phi$ is valid for each choice of $z_1,
\ldots, z_q$. This is done by the cycle $q \arrow{}{} q$, which
explores all combinations of $0$'s and $1$'s for the counters
$x_{p+\sum_{n=1}^{j}\pi_n}$, corresponding to $z_j$, for all $j \in
[1,q]$. To this end, we use the permutation matrix $\vec{M}$, which
consists of $\idmatrix_p$ and $q$ rotation blocks $\vec{M}_{\pi_j} \in
\set{0,1}^{\pi_j \times \pi_j}$ (Figure \ref{fig:mat-reduction}). The
valuation $\vec{v}_0$ ensures that the initial value of $x_{p+\sum_{n=1}^{j}\pi_n}$ is
$1$, for all $j\in[1,q]$, the other counters being $0$ initially
(Figure \ref{fig:mat-reduction}).

Intuitively, after $n$ iterations of the affine function
$(\vec{M},\vec{0})$, labeling the cycle $q \arrow{}{} q$ in $S_\Phi$,
we have $x_{p+\sum_{n=1}^{j}\pi_n}=1$ iff $n$ is a multiple of $\pi_j$. This fact
guarantees that all combinations of $0$'s and $1$'s for
$z_1,\ldots,z_q$ have been visited in $\Pi_{j=1}^q\pi_j$ iterations of
the cycle. The guard $g_1$, labeling the cycle, tests that, at each
iteration, the formula $\Psi$ is satisfied, using a standard encoding
of the CNF formula $\Psi$ as a linear system. Namely, each literal
$y_i$ (resp. $\neg y_i$) is encoded as the term $x_i$ (resp. $1 -
x_i$), and each $z_j$ (resp. $\neg z_j$) is encoded as
$x_{p+\sum_{n=1}^j \pi_n}$ (resp. $1 - x_{p+\sum_{n=1}^j
  \pi_n}$). Each disjunctive clause translates to a linear inequality
asking that the sum of the terms encoding the literals be greater or
equal to $1$, which forces at least one literal in each clause to be
true. For instance, the clause $y_1 \vee \neg z_2$ is encoded as $x_1
+ (1 -x_{p+\pi_1 + \pi_2}) \geq 1$ in $g_1$. Finally, the guard $g_2$
simply checks that $x_{\pi_1} = \ldots = x_{\pi_1 + \ldots + \pi_q} =
1$, ensuring that the loop has been iterated sufficiently many
times. This allows us to deduce the following result.

\begin{lemma}\label{lemma:reach-hardness}
$\Reach(\FlatMonoidAffineSystems)$ is $\Sigma_2^P$-hard.
\end{lemma}

\section{Bounding the Number of Cycle Iterations}
%%%%%%%%%%%%%%%%%%%%%%%%%%%%%%%%%%%%%%%%%%%%%%%%%%%%%%%%%%%%%%%%%%%%%%%%%%%%%%%%

In this section we prove a crucial property of counter systems from
the $\FlatMonoidAffineSystems$ class, namely that there exists a
polynomial function $\poly{x}$ such that, for each run $\rho$ starting
at $\aconf_0$ of the
considered counter system, there exists another run $\rho'$ starting
at $\aconf_0$, using the
same transition rules as $\rho$, in exactly the same order, and which
iterates each simple cycle at most
$2^{\poly{\sizeof{S}+\sizeof{\aconf_0}}}$ times. 

% In this section we give the first upper bound, for the reachability
% problem. This upper bound matches the lower bound from Section
% \ref{sec:lower-bounds}, showing that the class
% $\Reach(\FlatMonoidAffineSystems)$, of reachability problems for flat
% affine counter systems, with the finite monoid condition, is
% $\Sigma_2^p$-complete. This result improves the \textsc{4EXPtime}
% upper bound from Table \ref{tab:known-results}. The crux of the proof
% is to show the existence of a polynomial $\poly{x}$ such that, for
% each run $\rho$ in the (flat affine) counter system $S$, there exists
% another run $\rho'$, using the same transition rules as $\rho$, in
% exactly the same order, and which iterates each simple cycle
% $2^{\poly{\sizeof{S}}}$ times. Then we can use a polynomial-time
% bounded nondeterministic Turing machine that~\begin{inparaenum}[(i)]
% \item guesses the numbers of cycle iterations and
% %
% \item computes the effect of the updates along the guessed run, with an
% %
% \item \textsc{co-NP} oracle to check that no guard has been violated.
% \end{inparaenum}
% This gives us an $\textsc{NP}^{\textsc{co-NP}}$ algorithm for
% $\Reach(\FlatMonoidAffineSystems)$, which then lies in $\Sigma_2^p$

In the rest of this section, we fix a flat affine counter system $S =
\tuple{\States,\Counters_n,\Edges,\lab}$, with the finite monoid
property. We recall here that the set of runs of a flat counter system
can be captured by a finite (though exponential) number of \emph{path
  schemas} \cite{demri-complexity-13}. Formally, a path schema is a
non-empty finite sequence $P \equiv u_1 \ldots u_N$, where $u_i$ is
either a transition rule from $\Edges$ or a simple
cycle, such that \begin{inparaenum}[(i)]
\item $u_1, \ldots, u_N$ are pairwise distinct, 
\item $u_N$ is a simple cycle, called \emph{terminal}, and
\item $\target{u_i}=\source{u_{i+1}}$, for all $i \in [1,N-1]$. 
\end{inparaenum} 
All simple cycles on $P$, except for $u_N$, are called
\emph{nonterminal}. We use then the following notations:
$\lengthof{P}$ for $N$, $P[i]$ for $u_i$ with $i \in [1,N]$, and
$\sizeof{P}$  is the sum of the sizes of
all transition rules occurring in $P$.

Intuitively a path schema $P$ represents a set of infinite paths
obtained by iterating the non-terminal cycles a certain number of
times. We can hence represent such a path by its associated path
schema and an iteration vector. Formally, an \emph{iterated path
  schema} is a pair $\tuple{P,\vec{m}}$, such that $P$ is a path
schema, and $\vec{m} \in \nat^{\lengthof{P}-1}$ is a vector, where for
all $i \in [1,\lengthof{P}-1]$, $\vec{m}[i] \geq 1$ and $\vec{m}[i]>1$
implies that $P[i]$ is a cycle. An iterated path schema defines a
unique infinite word over $\Edges$, denoted by
\(\wordof{P,\vec{m}}=P[1]^{\vec{m}[1]}P[2]^{\vec{m}[2]} ~\cdots~
P[\lengthof{P}-1]^{\vec{m}[\lengthof{P}-1]}P[\lengthof{P}]^\omega\). We
recall the following result:
\begin{lemma}\cite{demri-taming-15}\label{lem:iterated-path-schema}
  Let $S$ be a flat affine counter system. Then:
  \begin{compactenum}
  \item\label{it1:iterated-path-schema} the length and the size of a
    path schema of $S$ are polynomial in $\sizeof{S}$.
  \item\label{it2:iterated-path-schema} for any run $\arun$ of $S$,
    there exists an iterated path schema $\tuple{P,\vec{m}}$
    such that $\wordof{\arun}=\wordof{P,\vec{m}}$.
  \end{compactenum}
\end{lemma}
For a run $\arun$, we consider the set $\ips{\arun} =
\set{\tuple{P,\vec{m}} \mid
  \wordof{\arun}=\wordof{P,\vec{m}}}$. Observe that
$\ips{\arun}\neq\emptyset$ for any run $\arun$ of $S$, due to Lemma
\ref{lem:iterated-path-schema} (\ref{it2:iterated-path-schema}).
Moreover, as a consequence of Lemma \ref{lem:iterated-path-schema}
(\ref{it1:iterated-path-schema}), the number of path schemas is
bounded by a simple exponential in the size of $S$. 

%% As an example, $P \equiv \astate_0 \arrow{\delta_0}{}\astate_0,
%% \astate_0 \arrow{\delta_1}{} \astate_1, \astate_1 \arrow{\delta_2}{}
%% \astate_1, \astate_1 \arrow{\delta_3}{} \astate_2, \astate_2
%% \arrow{\delta_4}{} \astate_2, \astate_2 \arrow{\delta_5}{} \astate_3,
%% \astate_3 \arrow{\delta_6}{} \astate_3$ is a path schema of the
%% counter system depicted in Figure \ref{fig:flat-cm}.

% \[P \equiv \astate_0 \arrow{\delta_1}{} \astate_1, \astate_1 \arrow{\delta_2}{} \astate_1, 
% \astate_1 \arrow{\delta_3}{} \astate_2, \astate_2 \arrow{\delta_4}{} \astate_2, 
% \astate_2 \arrow{\delta_5}{} \astate_3, \astate_3 \arrow{\delta_6}{} \astate_3
% \enspace.\] 

We fix a run $\arun$ of $S$ starting at $\aconf_0$, and $\tuple{P,\vec{m}} \in \ips{\arun}$
an iterated path schema corresponding to $\arun$. We consider a simple
cycle $c=\anedge_0 \ldots \anedge_{k-1}$ of $P$, whose transition
rules are $\anedge_i=\tuple{\astate_i, \vec{C}_i\cdot{x}\leq\vec{d}_i,
  (\vec{A}_i,\vec{b}_i), \astate_{i+1}}$, for all $i \in [0,k-1]$, and
$\astate_{k}=\astate_0$. Let $f_c = (\vec{A}_c,\vec{b}_c)$ be the
update of the entire cycle $c$, where $\vec{A}_c = \vec{A}_{k-1}
~\cdots~ \vec{A}_1 \cdot \vec{A}_0$, denoted $\prod_{i=k-1}^{0}
\vec{A}_i$, and $\vec{b}_c = \sum_{i=0}^{k-1} \prod_{j=k-1}^{i+1}
\vec{A}_j \cdot \vec{b}_i$. Since $S$ has the finite monoid property,
the set $\mathcal{M}_c = \set{\vec{A}_c^i \mid i \in \nat}$ is
finite. Then there exists two integer constants $\alpha, \beta \in
\nat$, such that $0 \leq \alpha+\beta \leq \card{\mathcal{M}_c}+1$,
and $\vec{A}_c^{\alpha} = \vec{A}_c^{\alpha+\beta}$. Observe that, in
this case, we have \(\mathcal{M}_c = \set{\vec{A}_c^0, \ldots,
  \vec{A}_c^\alpha, \ldots, \vec{A}_c^{\alpha+\beta-1}}\).

Our goal is to exhibit another run $\arun'$ of $S$ and an iterated
path schema $\tuple{P,\vec{m}'} \in \ips{\arun'}$, such that
$\card{\vec{m}'}_\infty \leq 2^{\poly{\sizeof{S}+\sizeof{\aconf_0}}}$, for a polynomial
function $\poly{x}$. Because $c=\anedge_0 \ldots \anedge_{k-1}$ is a
simple cycle of $P$ and $\tuple{P,\vec{m}} \in \ips{\arun}$, there
exists a (possibly infinite) subsequence of $\rho$, let us call it
\(\theta=(\astate_0,\vec{v}_0) \arrow{\tau_0}{} (\astate_1,\vec{v}_1)
\arrow{\tau_1}{} \ldots\) that iterates $c$, i.e.\ $\tau_i =
\anedge_{(i\!\!\mod\!  k)}$, for all $i\geq0$. In the following, we
call any subsequence of a run an \emph{execution}. 

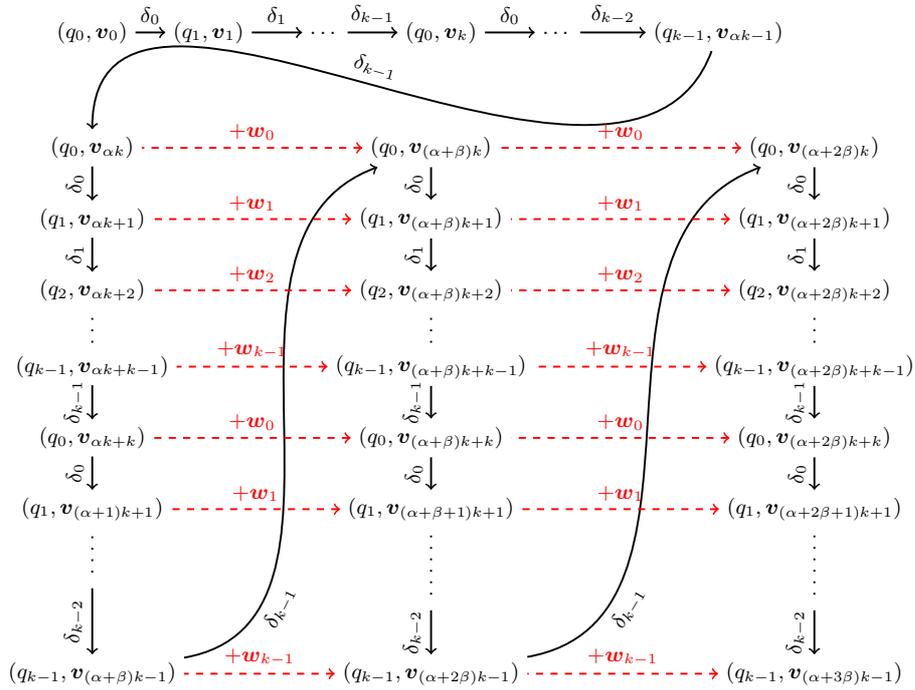
\begin{figure}[htb]
\vspace*{-2\baselineskip}
\begin{center}
\scalebox{0.9}{
\begin{tikzpicture}[node distance=1.7cm]
\node(0){$(\astate_0,\vec{v}_0)$};
\node(1)[right of=0]{$(\astate_1,\vec{v}_1)$} 
  edge [<-,thick] node[above] {$\anedge_0$} (0);
\node(2)[right of=1]{\ldots}
  edge [<-,thick] node[above] {$\anedge_1$} (1); ; 
\node(k)[,right of=2]{$(\astate_0,\vec{v}_k)$}
   edge [<-,thick] node[above] {$\anedge_{k-1}$} (2); 
\node(2bis)[right of=k]{\ldots}
  edge [<-,thick] node[above] {$\anedge_0$} (k); 
\node(alphak0)[right of=2bis,xshift=2em]{$(\astate_{k-1},\vec{v}_{{\alpha}k-1})$}
  edge [<-,thick] node[above] {$\anedge_{k-2}$} (2bis); 

\node(alphak)[below of=0]{$(\astate_0,\vec{v}_{{\alpha}k})$}
  edge [<-,thick,sloped,out=90,in=-110] node[above] {$\anedge_{k-1}$} (alphak0); 
\node(alphak1)[below of=alphak,yshift=2em]{$(\astate_1,\vec{v}_{{\alpha}k+1})$}
  edge [<-,thick,sloped] node [above] {$\anedge_{0}$} (alphak);
\node(alphak2)[below of=alphak1,yshift=2em]{$(\astate_2,\vec{v}_{{\alpha}k+2})$}
  edge [<-,thick,sloped] node [above] {$\anedge_{1}$} (alphak1);
\node(tmp)[below of=alphak2,yshift=2.5em]{}
  edge [loosely dotted,thick,sloped] node [above] {} (alphak2);
\node(alphak2bis)[below of=tmp,yshift=4.5em]{$(\astate_{k-1},\vec{v}_{{\alpha}k+k-1})$};
\node(alphakk)[below of=alphak2bis,yshift=2em]{$(\astate_0,\vec{v}_{{\alpha}k+k})$}
  edge [<-,thick,sloped] node [above] {$\anedge_{k-1}$} (alphak2bis);
\node(alphakk1)[below of=alphakk,yshift=2em]{$(\astate_1,\vec{v}_{({\alpha}+1)k+1})$}
  edge [<-,thick,sloped] node [above] {$\anedge_{0}$} (alphakk);
\node(tmp2)[below of=alphakk1,yshift=2em]{}
  edge [loosely dotted,thick,sloped] node [above] {} (alphakk1);
\node(alphakk3)[below of=tmp2,yshift=1em]{$(\astate_{k-1},\vec{v}_{({\alpha}+{\beta})k-1})$}
  edge [<-,thick,sloped] node [above] {$\anedge_{k-2}$} (tmp2);

\node(alphak-2)[right of=alphak,xshift=10em]{$(\astate_0,\vec{v}_{({\alpha}+{\beta})k})$}
  edge [<-,thick,sloped,in=10,out=200] node[pos=0.8,below] {$\anedge_{k-1}$}
  (alphakk3)
edge [<-,thick,sloped,dashed,red] node[above] {$+\vec{w}_0$}
  (alphak); 
\node(alphak1-2)[below of=alphak-2,yshift=2em]{$(\astate_1,\vec{v}_{({\alpha}+{\beta})k+1})$}
  edge [<-,thick,sloped] node [above] {$\anedge_{0}$} (alphak-2)
  edge [<-,thick,sloped,dashed,red] node[above] {$+\vec{w}_1$}
  (alphak1); 
\node(alphak2-2)[below of=alphak1-2,yshift=2em]{$(\astate_2,\vec{v}_{({\alpha}+{\beta})k+2})$}
  edge [<-,thick,sloped] node [above] {$\anedge_{1}$} (alphak1-2)
edge [<-,thick,sloped,dashed,red] node[above] {$+\vec{w}_2$}
  (alphak2);
\node(tmp-2)[below of=alphak2-2,yshift=2.5em]{}
  edge [loosely dotted,thick,sloped] node [above] {} (alphak2-2);
\node(alphak2bis-2)[below
of=tmp-2,yshift=4.5em]{$(\astate_{k-1},\vec{v}_{({\alpha}+{\beta})k+k-1})$}
edge [<-,thick,sloped,dashed,red] node[above] {$+\vec{w}_{k-1}$}
  (alphak2bis);
\node(alphakk-2)[below of=alphak2bis-2,yshift=2em]{$(\astate_0,\vec{v}_{({\alpha}+{\beta})k+k})$}
  edge [<-,thick,sloped] node [above] {$\anedge_{k-1}$} (alphak2bis-2)
edge [<-,thick,sloped,dashed,red] node[above] {$+\vec{w}_0$}
  (alphakk);
\node(alphakk1-2)[below of=alphakk-2,yshift=2em]{$(\astate_1,\vec{v}_{({\alpha}+{\beta}+1)k+1})$}
  edge [<-,thick,sloped] node [above] {$\anedge_{0}$} (alphakk-2)
edge [<-,thick,sloped,dashed,red] node[above] {$+\vec{w}_1$}
  (alphakk1);
\node(tmp2-2)[below of=alphakk1-2,yshift=1em]{}
  edge [loosely dotted,thick,sloped] node [above] {} (alphakk1-2);
\node(alphakk3-2)[below of=tmp2-2,yshift=2em]{$(\astate_{k-1},\vec{v}_{({\alpha}+2{\beta})k-1})$}
  edge [<-,thick,sloped] node [above] {$\anedge_{k-2}$} (tmp2-2)
edge [<-,thick,sloped,dashed,red] node[above] {$+\vec{w}_{k-1}$}
  (alphakk3);

\node(alphak-3)[right of=alphak-2,xshift=12em]{$(\astate_0,\vec{v}_{({\alpha}+2{\beta})k})$}
  edge [<-,thick,sloped,in=10,out=200] node[pos=0.8,below] {$\anedge_{k-1}$}
  (alphakk3-2)
edge [<-,thick,sloped,dashed,red] node[above] {$+\vec{w}_0$}
  (alphak-2); 
\node(alphak1-3)[below of=alphak-3,yshift=2em]{$(\astate_1,\vec{v}_{({\alpha}+2{\beta})k+1})$}
  edge [<-,thick,sloped] node [above] {$\anedge_{0}$} (alphak-3)
  edge [<-,thick,sloped,dashed,red] node[above] {$+\vec{w}_1$}
  (alphak1-2); 
\node(alphak2-3)[below of=alphak1-3,yshift=2em]{$(\astate_2,\vec{v}_{({\alpha}+2{\beta})k+2})$}
  edge [<-,thick,sloped] node [above] {$\anedge_{1}$} (alphak1-3)
edge [<-,thick,sloped,dashed,red] node[above] {$+\vec{w}_2$}
  (alphak2-2);
\node(tmp-3)[below of=alphak2-3,yshift=2.5em]{}
  edge [loosely dotted,thick,sloped] node [above] {} (alphak2-3);
\node(alphak2bis-3)[below
of=tmp-3,yshift=4.5em]{$(\astate_{k-1},\vec{v}_{({\alpha}+2{\beta})k+k-1})$}
edge [<-,thick,sloped,dashed,red] node[above] {$+\vec{w}_{k-1}$}
  (alphak2bis-2);
\node(alphakk-3)[below of=alphak2bis-3,yshift=2em]{$(\astate_0,\vec{v}_{({\alpha}+2{\beta})k+k})$}
  edge [<-,thick,sloped] node [above] {$\anedge_{k-1}$} (alphak2bis-3)
edge [<-,thick,sloped,dashed,red] node[above] {$+\vec{w}_0$}
  (alphakk-2);
\node(alphakk1-3)[below of=alphakk-3,yshift=2em]{$(\astate_1,\vec{v}_{({\alpha}+2{\beta}+1)k+1})$}
  edge [<-,thick,sloped] node [above] {$\anedge_{0}$} (alphakk-3)
edge [<-,thick,sloped,dashed,red] node[above] {$+\vec{w}_1$}
  (alphakk1-2);
\node(tmp2-3)[below of=alphakk1-3,yshift=1em]{}
  edge [loosely dotted,thick,sloped] node [above] {} (alphakk1-3);
\node(alphakk3-3)[below of=tmp2-3,yshift=2em]{$(\astate_{k-1},\vec{v}_{({\alpha}+3{\beta})k-1})$}
  edge [<-,thick,sloped] node [above] {$\anedge_{k-2}$} (tmp2-3)
edge [<-,thick,sloped,dashed,red] node[above] {$+\vec{w}_{k-1}$}
  (alphakk3-2);
\end{tikzpicture}}
\end{center}
\caption{Behavior of an execution which iterates ${\alpha}+3{\beta}$
  times the cycle $c=\anedge_0 \ldots \anedge_{k-1}$}
\label{fig:iterating-cycle}
\vspace*{-\baselineskip}
\end{figure}

The main intuition now is that $\theta$ can be decomposed into a
prefix of length $(\alpha+\beta)k$ and $k$ infinite sequences of
translations along some effectively computable vectors $\vec{w}_0,
\ldots, \vec{w}_{k-1}$. More precisely, all valuations $\vec{v}_i$ of
$\theta$, for $i \geq (\alpha+\beta)k$, that are situated at distance
$\beta k$ one from another, differ by exactly the same vector. 
%
%% We show that the sequence $\vec{v}_0,\vec{v}_1,\ldots$ of counter
%% valuations in $\theta$ can be described as follows: for all $i \geq
%% (\alpha+\beta)k$, each $\vec{v}_i$ is obtained by iteratively applying
%% a translation along some effectively computable vectors $\vec{w}_0,
%% \ldots, \vec{w}_{k-1}$ from one of the vectors $\vec{v}_{\alpha
%%   k},\ldots,\vec{v}_{(\alpha+\beta) k -1}$.
%
We refer to Figure \ref{fig:iterating-cycle} for an illustration of
this idea.

\begin{lemma}\label{lem:cycle-translation}
  Given an execution \((\astate_0,\vec{v}_0) \arrow{\delta_0}{} \ldots
  \arrow{\delta_{k-1}}{} (\astate_k,\vec{v}_k) \arrow{\delta_0}{}
  \ldots\) of $S$ that iterates a simple cycle $c=\anedge_0 \ldots
  \anedge_{k-1}$, there exist $\vec{w}_0, \ldots, \vec{w}_{k-1} \in
  \zed^n$, such that
  \(\vec{v}_{(\alpha+p\beta+r)k+q} = \vec{v}_{(\alpha+r)k+q} + p\cdot\vec{w}_q, 
  \text{ for all } p \geq 0, r \in [0,\beta-1] \text{ and } q \in [0,k-1]\), 
  where $f_c=(\vec{A}_c,\vec{b}_c)$ is the update of $c$ and
  $\alpha,\beta\geq0$ are such that $\vec{A}_c^\alpha =
  \vec{A}_c^{\alpha+\beta}$.
\end{lemma}

We distinguish now the case when $c$ is a nonterminal cycle of $P$,
iterated finitely many times, from the case when $c$ is terminal, thus
iterated ad infinitum. We consider first the case when $c$ is a
nonterminal cycle, taken a finite number of times. Viewing the
sequence of counter valuations, that occur during the unfolding of a
simple loop, as a set of translations by vectors $\vec{w}_0, \ldots,
\vec{w}_{k-1}$, prefixed by an initial sequence, allows us to reduce
the problem of checking the validity of the guards along this sequence
to checking the guards only in the beginning and in the end of each
translation by $\vec{w}_q$, for $q \in [0,k-1]$. This is possible
because each guard in the loop is defined by a convex vector set
$\set{\vec{v}\in\zed^n \mid \vec{C}\cdot\vec{v}\leq\vec{d}}$, for a
matrix $\vec{C} \in \zed^{m\times n}$ and a vector $\vec{d} \in
\zed^m$, thus a sequence of vectors produced by a translation cannot
exit and then re-enter the same guard, later on. This crucial
observation, needed to prove the upper bound, is formalized below.

We consider the relaxed transition relation $\leadsto \subseteq
(\States \times \zed^n) \times \Edges \times (\States \times \zed^n)$,
defined as $(\astate,\vec{v}) \overset{\anedge}{\leadsto}
(\astate',\vec{v}')$ iff $\source{\anedge}=\astate$,
$\vec{v}'=\update{\anedge}(\vec{v})$ and $\target{\anedge}=\astate'$.
Hence, $\leadsto$ allows to move from one configuration to another as
in $\arrow{}{}$, but without testing the guards. In the following, we
fix a sequence of configurations \(\anexec'=(\astate_0,\vec{v}_0)
\overset{\scriptscriptstyle\tau_0}{\leadsto} (\astate_1,\vec{v}_1)
\overset{\scriptscriptstyle\tau_1}{\leadsto} \ldots\) called a
\emph{pseudo-execution}. We assume, moreover, that $\anexec'$ iterates
the simple cycle $c = \anedge_0,\ldots,\anedge_{k-1}$ a finite number
of times, i.e.\ $\tau_i \equiv \anedge_{i\!\!  \mod\!  k}$, for all
$i\geq0$. To check whether $\anexec'$ is a real execution, it is
enough to check the guards in the first $\alpha+\beta+1$ and the last
$\beta$ iterations of the cycle, as shown by the following lemma:

\begin{lemma}\label{lem:convexity}
  For any $m > (\alpha+\beta+1)k$, given a finite pseudo-execution
  \((\astate_0,\vec{v}_0) \overset{\scriptscriptstyle\tau_0}{\leadsto}
  \ldots \overset{\scriptscriptstyle\tau_{m-1}}{\leadsto}
  (\astate_m,\vec{v}_m)\) of $S$, that iterates a nonterminal simple
  cycle $c=\anedge_0 \ldots \anedge_{k-1}$, \((\astate_0,\vec{v}_0)
  \arrow{\tau_0}{} \ldots \arrow{\tau_{m-1}}{} (\astate_m,\vec{v}_m)\)
  is an execution of $S$ iff $\vec{v}_i \models \guard{\tau_i}$, for
  all $i \in [0,(\alpha+\beta+1)k-1] \cup [m-\beta k,m-1]$.
\end{lemma}

The next step is to show that if a cycle is iterated $\ell$ times with
$\ell= \alpha +\beta+p\beta+r$ for some $p>0$ and $r\in[0,\beta-1]$,
starting with values $\vec{v} \in \zed^n$, then
$[\vec{v}[1],\ldots,\vec{v}[n],p]^\top$ is the solution of a system of
inequations $\vec{M}_c\cdot[\vec{y};z]^\top\leq \vec{n}_c$, where
$[\vec{y};z]=[y_1,\ldots,y_n,z]$ is a vector of $n+1$ variables. The
bound on the number of iterations follows from the theorem below, by
proving that the sizes of the entries of $\vec{M}_c$ and $\vec{n}_c$
(in binary) are bounded by a polynomial in $\sizeof{S}$.

\begin{theorem}\label{thm:ineq-sol-bound}
  Given $\vec{A} \in \zed^{m \times n}$ and $\vec{b} \in \zed^m$, for
  $n\geq2$, the system $\vec{A}\cdot\vec{x} \leq \vec{b}$ has a
  solution in $\nat^n$ iff it has a solution such that
  $\card{\vec{x}}_\infty \leq m^{2n} \cdot \card{\vec{A}}_{\max}^n \cdot
  \card{\vec{b}}_\infty$.
\end{theorem}

We recall that $c=\anedge_0,\ldots,\anedge_{k-1}$, where
$\guard{\anedge_i}\equiv\vec{C}_i\cdot\vec{x}\leq\vec{d}_i$,
$\update{\anedge_i}\equiv(\vec{A}_i,\vec{b}_i)$, and that $f_c =
(\vec{A}_c,\vec{b}_c)$ is the affine function defining the update of
the entire cycle. For any $j>0$, we define
$\vec{b}_c^j=\Sigma_{i=0}^{j-1}\vec{A}_c^i \cdot \vec{b}_c$, hence
$f^\ell_c = (\vec{A}_c^\ell,\vec{b}_c^\ell)$ is the update
corresponding to $\ell$ iterations of the cycle for a fixed integer
constant $\ell>0$. The following set of inequalities express that all
guards are satisfied within the $\ell$-th iteration of the cycle
starting at $\vec{v} \in \zed^n$:
\[
\left\{\vec{C}_p \cdot (\prod_{i=p-1}^0 \vec{A}_i \cdot \vec{A}_c^{\ell-1}\cdot\vec{v} + (\sum_{i=0}^{p-1} \prod_{j=p-1}^{i+1} \vec{A}_j) \cdot \vec{b}_j) \leq \vec{d}_p\right\}_{p=0}^{k-1}
\]

%% \[\begin{array}{rcl}
%% \vec{C}_0 \cdot (\vec{A}_c^{\ell-1}\cdot\vec{v} + \vec{b}_c^{\ell-1}) & \leq & \vec{d}_0 \\
%% \vec{C}_1 \cdot (\vec{A}_0 \cdot \vec{A}_c^{\ell-1}\cdot\vec{v} + \vec{A}_0\cdot\vec{b}_c^{\ell-1} + \vec{b}_0) & \leq & \vec{d}_1 \\
%% & \cdots & \\
%% \vec{C}_{k-1} \cdot (\prod_{i=k-2}^0 \vec{A}_i \cdot \vec{A}_c^{\ell-1}\cdot\vec{v} + (\sum_{i=0}^{k-2} \prod_{j=k-2}^{i+1} \vec{A}_j) \cdot \vec{b}_j) 
%% & \leq & \vec{d}_{k-1}
%% \end{array}\]
%% In the sequel, we use the following notations: 
%% \[\vec{M}_\ell = \left[ \begin{array}{c}\vec{C}_0 \cdot
%%                                      \vec{A}_c^{\ell-1}\\ \vdots \\ \vec{C}_{k-1} \cdot \prod_{i=k-2}^0 \vec{A}_i \cdot
%%   \vec{A}_c^{\ell-1}\end{array}\right] ~ \vec{n}_\ell =
%% \left [\begin{array}{c} \vec{d}_0-\vec{C}_0\cdot\vec{b}_c^{\ell-1} \\
%%          \vdots \\
%%   \vec{d}_{k-1}-\vec{C}_{k-1} \cdot (\sum_{i=0}^{k-2}
%%   \prod_{j=k-2}^{i+1} \vec{A}_j) \cdot \vec{b}_j \end{array} \right]\] 
%% we can write this system as $\vec{M}_\ell\cdot\vec{x}\leq\vec{n}_\ell$. With this
%% notation, the validity of the guards during the first $\ell$
%% iterations of the cycle are defined by the linear system below:
%% \[\left[ \begin{array}{c} \vec{M}_1\\ \vdots \\
%%           \vec{M}_\ell \end{array} \right] \cdot\vec{x} \leq
%% \left[ \begin{array}{c} \vec{n}_1\\ \vdots \\ \vec{n}_\ell\end{array} \right]\]
For technical reasons that will be made clear next, we do not need to
consider the case when the loop is iterated less than
$\alpha+2\beta+1$ times. We know, from Lemma \ref{lem:convexity}, that
checking whether a given cycle $c$ can be iterated $\ell >
\alpha+2\beta+1$ times from $\vec{v}$, reduces to checking the
validity of the guards during the first $\alpha+\beta+1$ and the last
$\beta$ iterations only. This condition is encoded by the union of the
linear inequality systems below:
\[
\left[ \begin{array}{c} \vec{M}_1 \\[-1mm] \ldots \\[-1mm]
         \vec{M}_{\alpha+\beta+1} \end{array} \right] \cdot \vec{v}
     \leq  \left[ \begin{array}{c} \vec{n}_1 \\[-1mm] \ldots \\[-1mm]
                    \vec{n}_{\alpha+\beta+1} \end{array} \right]
\hspace*{1cm} 
\left[ \begin{array}{c} \vec{M}_1 \\[-1mm] \ldots \\[-1mm]
         \vec{M}_{\beta} \end{array} \right] \cdot f^{\ell-\beta}_c(\vec{v})
     \leq  \left[ \begin{array}{c} \vec{n}_1 \\[-1mm] \ldots \\[-1mm]
                    \vec{n}_{\beta} \end{array} \right]
\]
Since we assumed that $\ell > \alpha+2\beta+1$, it follows that
$\ell-\beta = \alpha+p\beta+r$ for some $p>0$ and $r\in[0,\beta-1]$,
thus $f^{\ell-\beta}_c(\vec{v}) =
f^{\alpha+r}_c(\vec{v})+p\cdot\vec{w}_0 =
\vec{A}_c^{\alpha+r}\cdot\vec{v}+\vec{b}_c^{\alpha+r}+p\cdot\vec{w}_0$,
by Lemma \ref{lem:cycle-translation}. Then, for any finite execution
starting with $\vec{v}$, and consisting of $\alpha+p\beta+r$
iterations of $c$, we have that the column vector
$[\vec{v}[1],\ldots,\vec{v}[n],p]^\top$ is a solution of the linear
system $\vec{M}_{c,r} \cdot [\vec{y};z]^\top \leq \vec{n}_{c,r}$,
where:
\[
\vec{M}_{c,r} = \left[ \begin{array}{ccc}
\vec{M}_1  & & \vec{0}\\[-1mm]
& \ldots & \\[-1mm]
\vec{M}_{\alpha+\beta+1}  & & \vec{0}\\
 \vec{M}_1 \cdot \vec{A}_c^{\alpha+r} && \vec{M}_1 \cdot \vec{w}_0 \\[-1mm]
& \ldots & \\[-1mm]
\vec{M}_\beta \cdot \vec{A}_c^{\alpha+r} && \vec{M}_\beta \cdot \vec{w}_0 \end{array}\right] 
\hspace*{1cm}
\vec{n}_{c,r} = \left[ \begin{array}{c} 
\vec{n}_1 \\[-1mm] \ldots \\[-1mm]
\vec{n}_{\alpha+\beta+1} \\
\vec{n}_1 - \vec{M}_1 \cdot \vec{b}_c^{\alpha+r} \\[-1mm]
\ldots \\[-1mm]
\vec{n}_\beta - \vec{M}_\beta \cdot \vec{b}_c^{\alpha+r} \end{array}\right]
\]

%% Moreover, by the same argument, we obtain
%% $\vec{x}'=f_c^\ell(\vec{x})=\vec{A}_c^{\alpha+r}\cdot\vec{x}+\vec{b}_c^{\alpha+r}+(y+1)\cdot\vec{w}_0$
%% (this equality can also be written as a linear system of
%% inequalities). Then, for each fixed constant $r \in [0,\beta-1]$, the
%% effect of $\alpha+y\beta+r$ iterations of the simple cycle $c$ can be
%% encoded by a linear inequality system with variables
%% $[\vec{x};\vec{x}';y]^\top$, obtained by juxtaposing the systems
%% above.

We now consider the case when the simple cycle $c=\anedge_0 \ldots
\anedge_{k-1}$ is terminal and let $\vec{w}_0, \ldots, \vec{w}_{k-1}
\in \zed^n$ be the vectors from Lemma \ref{lem:cycle-translation}. We
say that $c$ is \emph{infinitely iterable} iff for all $i \in
[0,k-1]$, we have $\vec{C}_i\cdot\vec{w}_i \leq 0$. Since
$\vec{w}_0,\ldots,\vec{w}_{k-1}$ are effectively computable
vectors\footnote{They are defined in the proof of Lemma
  \ref{lem:cycle-translation}.}, this condition is effective. The next
lemma reduces the existence of an infinite iteration of the cycle to
the existence of an integer solution of a linear inequation system.

\begin{lemma}\label{lem:omega}
  Given an infinite pseudo-execution \((\astate_0,\vec{v}_0)
  \overset{\tau_0}{\leadsto} (\astate_1,\vec{v}_1)
  \overset{\tau_1}{\leadsto} \ldots\) of $S$, that iterates a terminal
  simple cycle $c=\anedge_0 \ldots \anedge_{k-1}$,
  \((\astate_0,\vec{v}_0) \arrow{\tau_0}{} (\astate_1,\vec{v}_1)
  \arrow{\tau_1}{} \ldots\) is an infinite execution of $S$ iff $c$ is
  infinitely iterable and $\vec{v}_i \models \guard{\tau_i}$, for all
  $i \in [0,(\alpha+\beta+1)k-1]$.
\end{lemma}
As a consequence, for an infinitely iterable cycle $c$, the existence
of an execution that iterates $c$ infinitely often is captured by the
linear system $\vec{M}_{c,\omega} \cdot \vec{y} \leq
\vec{n}_{c,\omega}$, where $\vec{M}_{c,\omega}$ and
$\vec{n}_{c,\omega}$ are obtained by stacking the matrices $\vec{M}_1,
\ldots, \vec{M}_{\alpha+\beta+1}$ and vectors $\vec{n}_1, \ldots,
\vec{n}_{\alpha+\beta+1}$, respectively. 

We have now all the ingredients needed to bound the number of cycle
iterations within the runs of a flat affine counter system having the
finite monoid property. The argument used in the proof relies on the
result of Theorem \ref{thm:ineq-sol-bound}, namely that the size of a
minimal solution of a linear system of inequalities is polynomially
bounded in the maximum absolute value of its coefficients, and the
number of rows, and exponentially bounded in the number of
columns. Since the number of rows depends on the maximum size of the
monoids of the update matrices in the counter system, we use the
results of Appendix \ref{app:algebra}, namely that the size of a
finite monoid of a square matrix is simply exponential in the
dimension of that matrix. 

\begin{theorem}\label{thm:bounded-iteration}
  Given a flat affine counter system $S =
  \tuple{\States,\Counters_n,\Edges,\lab}$, with the finite monoid
  property, for any run \(\arun\) of $S$, starting in
  \((\astate_0,\vec{v}_0)\), and any iterated path schema
  $\tuple{P,\vec{m}} \in \ips{\arun}$, there exists a run \(\arun'\),
  starting in \((\astate_0,\vec{v}_0)\), and an iterated path schema
  $\tuple{P,\vec{m}'} \in \ips{\arun'}$, such that
  $\card{\vec{m}'}_\infty \leq
  2^{\poly{\sizeof{S}+\sizeof{\vec{v}_0}}}$, for a polynomial function
  $\poly{x}$.
\end{theorem}

\section{The Complexities of Decision Problems for $\FlatMonoidAffineSystems$}

In this section, we will prove that the previous reasoning on iterated
path schemas allows us to deduce complexity bounds of the reachability
problems and of model-checking with $\PLTL$ and $\FO$ formulae for flat
counter systems with the finite monoid property.

\subsection{Reachability is $\Sigma_2^P$-complete }

In this section we give the first upper bound, for the reachability
problem. This upper bound matches the lower bound from Section
\ref{sec:lower-bounds}, showing that
$\Reach(\FlatMonoidAffineSystems)$ is $\Sigma_2^P$-complete. This
result improves the \textsc{4EXPTIME} upper bound from Table
\ref{tab:known-results}. The crux of the proof is based on the result
provided by Theorem \ref{thm:bounded-iteration} and it follows the
following reasoning: we use a polynomial-time bounded nondeterministic
Turing machine that guesses an iterated path schema and then a
\textsc{NP} oracle to check whether a guard has been violated.  This
gives us an $\textsc{NP}^{\textsc{NP}}$ algorithm for
$\Reach(\FlatMonoidAffineSystems)$, which then lies in
$\Sigma_2^P$. Theorem \ref{thm:bounded-iteration} ensures us the
soundness of the Algorithm and the correctness is provided by the fact
that if, in an iterated path schema, no guard is violated then it
corresponds necessarily to a run.

Let us now explain how our \textsc{NP} oracle works. The next lemma is
based on the fact that any power $\vec{A}^k$ of a finite monoid matrix
$\vec{A}$ can be computed in time polynomial in $\sizeof{\vec{A}}$ and
$\log_2k$, using matrix exponentiation by squaring. The reason is that
the value of an entry of any power of a finite monoid matrix $\vec{A}$
is bounded by an exponential in $\sizeof{\vec{A}}$, thus the size of
its binary representation is polynomially bounded by
$\sizeof{\vec{A}}$, and each step of the squaring algorithm takes
polynomial time (Lemma \ref{lem:finite-monoid-powers-bound} in Appendix \ref{app:algebra}).

% This result is now applied to find a $\Sigma_2^p$ upper bound for the
% reachability problem. The next lemma is based on the fact that any
% power $\vec{A}^k$ of a finite monoid matrix $\vec{A}$ can be computed
% in time polynomial in $\sizeof{\vec{A}}$ and $\log_2k$, using matrix
% exponentiation by squaring. The reason is that the value of an entry
% of any power of a finite monoid matrix $\vec{A}$ is bounded by an
% exponential in $\sizeof{\vec{A}}$, thus the size of its binary
% representation is polynomially bounded by $\sizeof{\vec{A}}$, and each
% step of the squaring algorithm takes polynomial time.

\begin{lemma}\label{lem:co-np-oracle}
  Given an iterated path schema $\tuple{P,\vec{m}}$ of a counter
  system with the finite monoid property $S$ and an initial
  configuration $\aconf_0$, checking whether there is no run $\rho$
  starting at $\aconf_0$ such that $\tuple{P,\vec{m}} \in
  \ips{\arun}$ is in \textsc{NP}.
\end{lemma}

The next theorem  gives the main result of this section.

\begin{theorem}\label{cor:reach-sigma-2-p}
 $\Reach(\FlatMonoidAffineSystems)$ is
  $\Sigma^P_2$-complete.
\end{theorem}

\subsection{$\PLTL$ Model Checking is $\Sigma_2^P$-complete}

For a $\PLTL$ formula $\phi$, its temporal depth $td(\phi)$ is defined
as the maximal nesting depth of temporal operators in $\phi$, and the
size of $\phi$ is its number of subformulae.  In \cite[Theorem
  4.1]{demri-taming-15}, the authors have proved a \emph{stuttering}
theorem for $\PLTL$ stating that if an $\omega$-word $w=w_1 w^M_2 w_3$
over the alphabet $2^{\AtProps}$ with $w_2 \neq \epsilon$ satisfies a
PLTL formula $\phi$ (i.e. $w,0 \modelspltl \phi$) and if $M \geq
2td(\phi) + 5$ then all $\omega$-words $w'=w_1 w^{M'}_2 w_3$ with $M'
\geq 2td(\phi) + 5$ are such that $w',0 \modelspltl \phi$. In other
words, to verify if an $\omega$-word with some repeated infix words
satisfies a PLTL formula it is enough to verify the property for the
$\omega$-word where each infix is repeated at most $2td(\phi)+5$
times. This allows to deduce that the model-checking of PLTL for flat
translating counter systems is \textsc{NP}-complete. We rewrite now in
our terminology the main proposition which leads to this result.

In the sequel we consider a flat counter system $S =
\tuple{\States,\Counters_n,\Edges,\lab}$ with the finite monoid
property. For a finite sequence of transitions $\delta_1 \ldots d_k$,
we denote by $\lab(\delta_1 \ldots d_k) =\lab(\source{\delta_1})
\ldots \lab(\source{\delta_k})$ the finite word labeling the sequence
with sets of atomic propositions. We lift this definition to iterated
path schemas $\tuple{P,\vec{m}}$ as
$\lab(P,\vec{m})=\lab(P[1])^{\vec{m}[1]}\lab(P[2])^{\vec{m}[2]}
~\cdots~
\lab(P[\lengthof{P}-1])^{\vec{m}[\lengthof{P}-1]}\linebreak[0]\lab(P[\lengthof{P}])^\omega$. Observe
that, for a run $\arun$ of a counter system, if $\tuple{P,\vec{m}} \in
\ips{\arun}$ is an iterated path schema, we have by definition of the
semantics of $\PLTL$ that $\arun \modelspltl \phi$ iff
$\lab(P,\vec{m}),0 \modelspltl \phi$ for all $\PLTL$ formulae
$\phi$. Moreover, for each $m \in \nat$, we define the function
$\xi_m$ mapping each vector $\vec{v} \in \nat^k$ to $\xi_m(\vec{v})
\in \nat^k$, where, for all $i \in [1,k]$:
\[
\xi_m(\vec{v})[i]= \begin{cases}
\vec{v}[i] \mbox{ if } \vec{v}[i] < m \\
m \mbox{ otherwise}
\end{cases}
\]
Let us now recall the main technical propositions established in
\cite{demri-taming-15}, which are a consequence of the stuttering
theorem for $\PLTL$ and of the result on the complexity of
model-checking ultimately periodic path with $\PLTL$ given in
\cite{markey-model-03}.

\begin{lemma}\label{lemma:pltl-path-schema}
Let $\tuple{P,\vec{m}}$ be an iterated path schema and $\phi$ a $\PLTL$
formula, then: 
\begin{compactenum}
\item \cite[Proposition 5.1]{demri-taming-15} $\lab(P,\vec{m}),0 \modelspltl \phi$ iff $\lab(P,\xi_{2
    td(\phi)+5}(\vec{m})),0 \modelspltl \phi$, 
\item \cite[Theorem 3.2]{markey-model-03} Given finite words $u$ and
  $v$, checking $uv^{\omega},0 \modelspltl \phi$ can be done in time
  polynomial in the sizes of $uv$ and $\phi$.
\end{compactenum}
\end{lemma}

We need furthermore a version of Theorem \ref{thm:bounded-iteration}
above, which ensures that given an iterated path schema and a $\PLTL$
formula $\phi$, we do not change the number of times a loop is
iterated if this one is less than $2.td(\phi)+5$. The proof of the
next result can in fact be deduced by adapting the proof of Theorem
\ref{thm:bounded-iteration} by unfolding the loop which are iterated
less than $2.td(\phi)+5$ for a given formula $\phi$.  As a consequence
of Lemma \ref{lemma:pltl-path-schema}, the new run $\rho'$, obtained
in the next lemma, is such that $\rho \modelspltl \phi$ iff $\rho'
\modelspltl \phi$ for the considered $\PLTL$ formula $\phi$.

\begin{lemma}\label{lem:bounded-iteration-pltl}
  For a run \(\arun\) of $S$ starting in
  \((\astate_0,\vec{v}_0)\), an iterated path schema
  $\tuple{P,\vec{m}} \in \ips{\arun}$ and a $\PLTL$
formula $\phi$, there exists a run \(\arun'\)
  starting in \((\astate_0,\vec{v}_0)\), and an iterated path schema
  $\tuple{P,\vec{m}'} \in \ips{\arun'}$, such that
  $\card{\vec{m}'}_\infty \leq
  2^{\poly{\sizeof{S}+\sizeof{\vec{v}_0}+td(\phi)}}$ for a polynomial
  $\poly{x}$ and $\xi_{2td(\phi)+5}(\vec{m})=\xi_{2td(\phi)+5}(\vec{m'})$.
\end{lemma}

We can now explain why the model-checking of flat counter systems with
the finite monoid property with $\PLTL$ formulae is in $\Sigma_2^P$.
Given a flat counter system $S$ with the finite monoid property, an
initial configuration $\aconf_0$, and a $\PLTL$ formula $\phi$, we
guess an iterated path schema $\tuple{P,\vec{m}}$ of polynomial size
in the size of $S$, $ \aconf_0$ and $\phi$ and we check whether
$\lab(P,\xi_{2 td(\phi)+5}(\vec{m})),0 \modelspltl \phi$. This check
can be done in polynomial time in the size of $P$ and $\phi$ thanks to
Lemma \ref{lemma:pltl-path-schema}. Finally, we use the \textsc{NP}
algorithm of Lemma \ref{lem:co-np-oracle} to verify that there exists
a run $\rho$ starting at $\aconf_0$, such that $\tuple{P,\vec{m}} \in
\ips{\rho}$. This gives us a $\Sigma_2^P$ algorithm whose correctness
is ensured by Lemma \ref{lem:bounded-iteration-pltl} and Lemma
\ref{lem:iterated-path-schema}. We deduce the $\Sigma_2^P$ hardness
from Lemma \ref{lemma:reach-hardness}, since we can encode
reachability of a state into a $\PLTL$ formula.

\begin{theorem}
\label{thm:model-checking-pltl}
$\MC_\PLTL(\FlatMonoidAffineSystems)$  is $\Sigma_2^P$-complete.
\end{theorem}

\subsection{$\FO$ Model Checking is \textsc{PSPACE}-complete}

For a $\FO$ formula $\phi$, its quantifier height $qh(\phi)$ is the
maximal nesting depth of its quantifiers, and the size of $\phi$ is
its number of subformulae. Similarly, as for the $\PLTL$ case, in
\cite[Theorem 6]{demri-complexity-13}, a stuttering theorem for $\FO$
is provided, which says that that two $\omega$-words $w=w_1 w^M_2 w_3$
and $w=w_1 w^{M'}_2 w_3$ with $w \neq \epsilon$ are indistinguishable
by a $\FO$ formula $\phi$ if $M$ and $M'$ are strictly bigger than
$2^{qh(\phi)+2}$. The main difference with $\PLTL$ is that this
provides an exponential bound in the maximum number of times an infix
of an $\omega$-word needs to be repeated to satisfy a $\FO$
formula. In the sequel we consider a flat counter system $S =
\tuple{\States,\Counters_n,\Edges,\lab}$ with the finite monoid
property and we reuse the notations introduced in the previous
section. The results of \cite{demri-complexity-13} can be
restated as follows.

\begin{lemma}\label{lem:fo-path-schema}
Given an iterated path schema $\tuple{P,\vec{m}}$ and a $\PLTL$
formula $\phi$, the following hold: 
\begin{compactenum}
\item \cite[Lemma 7]{demri-complexity-13} $\lab(P,\vec{m})
  \modelsfo\phi$ iff $\lab(P,\xi_{2^{qh(\phi)+2}}(\vec{m})) \modelsfo
  \phi$,
\item \cite[Theorem 9]{demri-complexity-13} Checking
  $\lab(P,\vec{m}),0 \modelsfo\phi$ can be done in space polynomial in
  the sizes of $\tuple{P,\vec{m}}$ and $\phi$.
\end{compactenum}
\end{lemma}

As for the $\PLTL$ case, this allows us to deduce a \textsc{NPSPACE}
algorithm for the model-checking problem of flat counter system with
the finite monoid property with $\FO$ formulae. Since the problem is
already \textsc{PSPACE}-hard for flat translating counter systems
\cite[Theorem 9]{demri-complexity-13}, we conclude by the following
theorem.

\begin{theorem}
\label{thm:model-checking-fo}
$\MC_\FO(\FlatMonoidAffineSystems)$  is \textsc{PSPACE}-complete.
\end{theorem}

\bibliographystyle{splncs03}
\bibliography{biblio}

\begin{thebibliography}{10}
\providecommand{\url}[1]{\texttt{#1}}
\providecommand{\urlprefix}{URL }

\bibitem{AroraBarakBook}
Arora, S., Barak, B.: Computational complexity: a modern approach. Cambridge
  University Press (2009)

\bibitem{Fast}
Bardin, S., Finkel, A., Petrucci, J.L.L.: Fast: Fast acceleration of symbolic
  transition systems. \url{http://tapas.labri.fr/trac/wiki/FASTer}

\bibitem{BlondinFinkelGoellerHaaseMcKenzie05}
Blondin, M., Finkel, A., G{\"{o}}ller, S., Haase, C., McKenzie, P.:
  Reachability in two-dimensional vector addition systems with states is
  pspace-complete. CoRR  abs/1412.4259 (2014),
  \url{http://arxiv.org/abs/1412.4259}

\bibitem{BoigelotPhD}
Boigelot, B.: Symbolic Methods for Exploring Infinite State Spaces. PhD, Univ.
  de Li\`ege (1999)

\bibitem{BIKLmcs14}
Bozga, M., Iosif, R., Konecn{\'{y}}, F.: Deciding conditional termination.
  Logical Methods in Computer Science  10(3) (2014)

\bibitem{Cav10}
Bozga, M., Iosif, R., Kone\v{c}n\'{y}, F.: Fast acceleration of ultimately
  periodic relations. In: CAV. LNCS, vol. 6174, pp. 227--242 (2010)

\bibitem{ComonJurski99}
Comon, H., Jurski, Y.: Timed automata and the theory of real numbers. In:
  {CONCUR} '99, Proceedings. pp. 242--257 (1999)

\bibitem{demri-complexity-13}
Demri, S., Dhar, A.K., Sangnier, A.: On the complexity of verifying regular
  properties on flat counter systems,. In: {ICALP}'13. LNCS, vol. 7966, pp.
  162--173 (2013)

\bibitem{demri-equivalence-rp14}
Demri, S., Dhar, A.K., Sangnier, A.: Equivalence between model-checking flat
  counter systems and presburger arithmetic. In: {RP}'14. LNCS, vol. 8762, pp.
  85--97. Springer (2014)

\bibitem{demri-taming-15}
Demri, S., Dhar, A.K., Sangnier, A.: Taming past {LTL} and flat counter
  systems. Inf. Comput.  242,  306--339 (2015)

\bibitem{demri-model-10}
Demri, S., Finkel, A., Goranko, V., van Drimmelen, G.: Model-checking {CTL}*
  over flat presburger counter systems. Journal of Applied Non-Classical Logics
   20(4),  313--344 (2010)

\bibitem{FinkelLeroux02}
Finkel, A., Leroux, J.: How to compose presburger-accelerations: Applications
  to broadcast protocols. In: FST TCS '02. pp. 145--156 (2002)

\bibitem{MonniauxGawlitza12}
Gawlitza, T.M., Monniaux, D.: Invariant generation through strategy iteration
  in succinctly represented control flow graphs. LMCS  8(3) (2012)

\bibitem{goller-model-icalp10}
G{\"{o}}ller, S., Haase, C., Ouaknine, J., Worrell, J.: Model checking succinct
  and parametric one-counter automata. In: {ICALP}'10. LNCS, vol. 6199, pp.
  575--586. Springer (2010)

\bibitem{IbarraGurari81}
Gurari, E.M., Ibarra, O.H.: The complexity of decision problems for finite-turn
  multicounter machines. J. Computer and System Sciences  22,  220--229 (1981)

\bibitem{haase-subclasses-14}
Haase, C.: Subclasses of {P}resburger arithmetic and the weak {EXP} hierarchy.
  In: {CSL-LICS} '14. pp. 47:1--47:10. {ACM} (2014)

\bibitem{Atva12}
Hojjat, H., Iosif, R., Kone{\v{c}}n{\'y}, F., Kuncak, V., R{\"u}mmer, P.:
  Accelerating Interpolants, pp. 187--202. Springer Berlin Heidelberg (2012)

\bibitem{Flata}
Konecny, F., Iosif, R., Bozga, M.: Flata: a verification toolset for counter
  machines. \url{http://nts.imag.fr/index.php/Flata} (2009)

\bibitem{kuhtz-weak-concur11}
Kuhtz, L., Finkbeiner, B.: Weak {K}ripke structures and {LTL}. In: {CONCUR}'11.
  LNCS, vol. 6901, pp. 419--433. Springer (2011)

\bibitem{LerouxSutre04}
Leroux, J., Sutre, G.: On Flatness for 2-Dimensional Vector Addition Systems
  with States, pp. 402--416 (2004)

\bibitem{Lipton76}
Lipton, R.J.: The reachability problem is exponential-space-hard. Tech.
  Rep.~62, Yale University, Department of Computer Science (1976)

\bibitem{markey-model-03}
Markey, N., Schnoebelen, P.: Model checking a path. In: {CONCUR}'03. LNCS, vol.
  2761, pp. 248--262 (2003)

\bibitem{Minsky67}
Minsky, M.: Computation: Finite and Infinite Machines. Prentice-Hall (1967)

\bibitem{Schrijver86}
Schrijver, A.: Theory of Linear and Integer Programming. Wiley \& Sons, Inc.
  (1986)

\bibitem{sistla-complexity-85}
Sistla, A., Clarke, E.: The complexity of propositional linear temporal logic.
  J.~ACM  32(3),  733--749 (1985)

\bibitem{stockmeyer-complexity-74}
Stockmeyer, L.J.: The complexity of decision problems in automata and logic.
  Ph.D. thesis, MIT (1974)

\end{thebibliography}

%%%%%%%%%%%%%%%%%%%%%%%%%%%%%%%%%%%%%%%%%%%%%%%%%%%%%%%%%%%%%%%%%%%%%%%%%%%%%%%%
\newpage
\appendix 
%%%%%%%%%%%%%%%%%%%%%%%%%%%%%%%%%%%%%%%%%%%%%%%%%%%%%%%%%%%%%%%%%%%%%%%%%%%%%%%%

%%%%%%%%%%%%%%%%%%%%%%%%%%%%%%%%%%%%%%%%%%%%%%%%%%%%%%%%%%%%%%%%%%%%%%%%%%%%%%%%
\section{Proofs of the Main Results}\label{app:proofs}
%%%%%%%%%%%%%%%%%%%%%%%%%%%%%%%%%%%%%%%%%%%%%%%%%%%%%%%%%%%%%%%%%%%%%%%%%%%%%%%%

%-------------------------------------------------------------------------------
\subsection{Proof of Lemma \ref{lemma:reach-hardness}}
%-------------------------------------------------------------------------------

\begin{proof}
It is not hard to check that $\Phi$ is valid if and only if $S_\Phi$
has a run that starts in $(\astate_0,\vec{v}_0)$. Moreover, $S_\Phi$ is flat and
the multiplicative monoid of the powers of $\vec{M}$ is finite. This
is because $\idmatrix_p$ and each rotation block $\vec{M}_{\pi_j}$
have a finite monoid. It remains only to show that the construction of
$S_\Phi$ is possible in time polynomial in the size of $\Phi$. The
only non-trivial point here is proving that $N=p+\sum_{j=1}^q \pi_j$ is
polynomial in $p$ and $q$. By the Prime Number Theorem we have that
$\pi_n \sim n \log n$, the asymptotic notation meaning that the error
approaches $0$ as $n$ increases. Then we have $N \sim p + \sum_{j=1}^q
j \log j \leq p + q \log q$ and we are done. \qed
\end{proof}

%-------------------------------------------------------------------------------
\subsection{Proof of Lemma \ref{lem:cycle-translation}}
%-------------------------------------------------------------------------------

As appetizer, the following lemma gives a property of the sequences of
vectors obtained by iteratively applying an affine function:

\begin{lemma}\label{lem:iterating-func}
Given an affine function $f = (\vec{A},\vec{b})$ and two
integers $\alpha,\beta\geq0$, such that $\vec{A}^\alpha =
\vec{A}^{\alpha+\beta}$, for all $\vec{v} \in \zed^n$, $r \in
    [0,\beta-1]$ and $p\in \nat$, we have: \[f^{\alpha+p\beta+r}(\vec{v}) =
    f^{\alpha+r}(\vec{v}) + p \cdot \vec{A}^{\alpha} \cdot
    f^{\beta}(\vec{0})\enspace.\]
\end{lemma}
\begin{proof}
We compute:
\[\begin{array}{rcl}
\afunc^{\alpha+\beta}(\vec{v}) & = &
                                           \vec{A}^{\alpha+\beta}
                                           \cdot \vec{v} +
                                           \vec{A}^{\alpha+\beta-1}
                                           \cdot \vec{b} + \vec{A}^{\alpha+\beta-2}
                                           \cdot \vec{b}  + \ldots
                                           +\vec{b}\\
  &= &\vec{A}^{\alpha} \cdot \vec{v} + \vec{A}^{\alpha-1}
       \cdot \vec{b} + \ldots +\vec{b} + 
                                           \vec{A}^{\alpha+\beta-1}
                                           \cdot \vec{b} + \vec{A}^{\alpha+\beta-2}
                                           \cdot \vec{b}  + \ldots +\vec{A}^{\alpha}
                                           \cdot \vec{b} \\
&=& \afunc^{\alpha}(\vec{v}) + 
                                           \vec{A}^{\alpha} \cdot
    (\vec{A}^{\beta-1}
                                           \cdot \vec{b} + \vec{A}^{\beta-2}
                                           \cdot \vec{b}  + \ldots
    +\vec{b} )\\
&=& \afunc^{\alpha}(\vec{v}) + 
                                           \vec{A}^{\alpha} \cdot
    (\afunc^{\beta}(\vec{0}))\enspace.
\end{array}\]
Hence, for all $p>0$:
$\afunc^{\alpha+p\beta}(\vec{v})=\afunc^{\alpha+\beta}(\afunc^{(p-1)\beta}(\vec{v}))
=\afunc^{\alpha+(p-1)\beta}(\vec{v})+\vec{A}^{\alpha}\cdot\afunc^{\beta}(\vec{0})$. Applying
this argument inductively, we obtain:
\(\afunc^{\alpha+p\beta}(\vec{v})=\afunc^{\alpha}(\vec{v})+p\cdot\vec{A}^{\alpha}\cdot
\afunc^{\beta}(\vec{0})\), for all $p \geq 0$. Finally for all $r
\in [0,\beta-1]$ and $p \in \nat$, we have:
\(f^{\alpha+p\beta+r}(\vec{v}) =f^{\alpha+p\beta}(f^r(\vec{v}))
=f^{\alpha+r}(\vec{v}) + p \cdot \vec{A}^{\alpha} \cdot
f^{\beta}(\vec{0})\).\qed
\end{proof}

\begin{proof} We compute: 
\[\begin{array}{rcl}
\vec{v}_{(\alpha+p\beta+r)k+q} & = & f_{q-1}( \ldots f_0(f_c^{\alpha+p\beta+r}(\vec{v}_0)) \ldots) \\
& = & f_{q-1}( \ldots f_0(f_c^{\alpha+r}(\vec{v}_0) + p \cdot \vec{A}_c^\alpha \cdot f_c^\beta(\vec{0})) \ldots) \text{, by Lemma \ref{lem:iterating-func}} \\
& = & f_{q-1}( \ldots f_0(f_c^{\alpha+r}(\vec{v}_0)) \ldots) + p \cdot \prod_{i=q-1}^0 \vec{A}_i \cdot \vec{A}_c^\alpha \cdot f_c^\beta(\vec{0}) \\
& = & \vec{v}_{(\alpha+r)k + q} + p \cdot \prod_{i=q-1}^0 \vec{A}_i \cdot \vec{A}_c^\alpha \cdot f_c^\beta(\vec{0})
\end{array}\]
By setting $\vec{w}_0 = \vec{A}_c^\alpha \cdot f_c^\beta(\vec{0})$ and
$\vec{w}_{i+1} = \vec{A}_i \cdot \vec{w}_i$, for all $i \in [0,k-2]$,
we obtain the result.\qed
\end{proof}

%-------------------------------------------------------------------------------
\subsection{Proof of Lemma \ref{lem:convexity}}
%-------------------------------------------------------------------------------

\begin{proof}
  ``$\Rightarrow$'' This direction is trivial. ``$\Leftarrow$'' It is
  sufficient to show that $\vec{v}_i \models \guard{\tau_i}$, for all
  $i \in [(\alpha+\beta)k, m-\beta k-1]$. The hypothesis ensures then
  that $\vec{v}_i \models \guard{\tau_i}$ for all $i \in [0,m-1]$,
  thus \((\astate_0,\vec{v}_0) \arrow{\tau_0}{} \ldots
  \arrow{\tau_{m-1}}{} (\astate_m,\vec{v}_m)\) is actually an
  execution. We assume that $(\alpha+\beta+1)k < m-\beta k$, otherwise
  $[0,(\alpha+\beta+1)k-1] \cup [m-\beta k,m-1] = [0,m-1]$ and there
  is nothing to prove. If $i \in [(\alpha+\beta+1)k, m-\beta k-1]$,
  there exist $p\geq1$, $r \in [0,\beta-1]$ and $q \in [0,k-1]$ such
  that $i=(\alpha+p\beta+r)k+q$. By Lemma \ref{lem:cycle-translation},
  there exists a vector $\vec{w}_q$ such that $\vec{v}_i =
  \vec{v}_{(\alpha+r)k+q} + p\cdot\vec{w}_q$. Then $0 \leq
  (\alpha+r)k+q < (\alpha+\beta+1)k$, thus $\vec{v}_{(\alpha+r)k+q}
  \models \guard{\tau_{(\alpha+r)k+q}}$, by the hypothesis. Let $p' =
  \lceil \frac{m-i}{\beta k} \rceil+p-1$. Since $i < m - \beta k$, we
  have $p' > p$, and we denote $j = (\alpha+p'\beta+r)k+q$. It is easy
  to check that $j \in [m-\beta k, m-1]$ and, by the hypothesis, we
  have $\vec{v}_j \models \guard{\tau_j}$. Since $\vec{v}_i =
  \vec{v}_{(\alpha+r)k+q} + p \cdot \vec{w}_q$ and $\vec{v}_j =
  \vec{v}_i + (p'-p) \cdot \vec{w}_q$, we obtain that $\vec{v}_i
  \models \guard{\tau_i}$, because $\tau_{(\alpha+r)k+q} = \tau_i =
  \tau_j = \delta_q$ and $\guard{\delta_q}$ is a convex polyhedron.\qed
\end{proof}

%-------------------------------------------------------------------------------
\subsection{Proof of Theorem \ref{thm:ineq-sol-bound}}
%-------------------------------------------------------------------------------

\proof{If $\vec{A}\cdot\vec{x} \leq \vec{b}$ has a solution in
  $\nat^n$, both sets $\set{\vec{x} \in \rat_{\geq0}^n \mid \vec{A}
    \cdot \vec{x} \leq \vec{b}}$ and $\set{\vec{x} \in \nat^n \mid
    \vec{A} \cdot \vec{x} \leq \vec{b}}$ are non-empty. Then there
  exists $\vec{a} \in \zed^{1 \times n}$ such
  that \begin{inparaenum}[(i)]
    \item\label{it:max-rat} $\max\set{\vec{a}\cdot\vec{x} \mid \vec{x}
      \in \rat_{\geq0}^n, \vec{A} \cdot \vec{x} \leq \vec{b}}$ and
    \item\label{it:max-int} $\max\set{\vec{a}\cdot\vec{x} \mid \vec{x}
      \in \nat^n, \vec{A} \cdot \vec{x} \leq \vec{b}}$
  \end{inparaenum}
  are both finite (take $\vec{a}$ to be the $i$-th row of $\vec{A}$
  and obtain $\vec{a}\vec{x} \leq \vec{b}[i]$). Let $\vec{z} \in
  \nat^n$ be an optimal solution of (\ref{it:max-int}), thus a
  solution of $\vec{A}\cdot\vec{x}\leq\vec{b}$. By \cite[Theorem
    17.2]{Schrijver86}, there exists an optimal solution $\vec{y} \in
  \rat_{\geq0}^n$ of (\ref{it:max-rat}), which is the solution of a
  subsystem of equations $\vec{A}'\cdot\vec{x}=\vec{b}'$, where
  $\vec{A}'$ and $\vec{b}'$ are obtained by deleting several rows from
  $\vec{A}$ and $\vec{b}$, respectively. Then either $\vec{y}$ is a
  vertex of the polyhedron $\vec{A}\cdot\vec{x}\leq\vec{b}$, in which
  case the solution of $\vec{A}'\cdot\vec{x}=\vec{b}'$ is unique, or a
  facet thereof, in which case all solutions of
  $\vec{A}'\cdot\vec{x}=\vec{b}'$ are optimal. Moreover, there exist
  vectors $\vec{c}_0, \vec{c}_1, \ldots, \vec{c}_t \in \rat^n$ such that
  $\set{\vec{x} \in \rat^n_{\geq0} \mid \vec{A}'\cdot\vec{x}=\vec{b}'}
  = \set{\vec{c}_0 + \sum_{i=1}^t \vec{c}_i \lambda_i \mid \lambda_1,
    \ldots, \lambda_t \in \rat_{\geq0}}$ and each non-zero component
  of $\vec{c}_0, \vec{c}_1, \ldots, \vec{c}_t$ is a quotient of
  subdeterminants of the matrix $[\vec{A}'; \vec{b}']$. Thus
  $\vec{c}_0$ is a solution of $\vec{A}'\cdot\vec{x}=\vec{b}'$ and we
  have $\card{\vec{c}_0}_\infty \leq \Delta$, where $\Delta$ is the
  maximum absolute value of the subdeterminants of
  $[\vec{A};\vec{b}]$. Again by \cite[Theorem 17.2]{Schrijver86},
  there exists an optimal solution $\vec{z}_0 \in \nat^n$ of
  (\ref{it:max-int}) such that $\card{\vec{z}_0-\vec{c}_0}_\infty \leq
  n\Delta$. By the triangle inequality, we obtain:
  \[\begin{array}{rcl}
  \card{\vec{z}_0}_\infty & \leq & \card{\vec{z}_0-\vec{c}_0}_\infty + \card{\vec{c}_0}_\infty \\
  & \leq & (n+1)\Delta
  \end{array}\]
  By Hadamard's inequality, we have that:
  \[\Delta ~\leq~ \card{\vec{A}[1]}_2 ~\cdots~ \card{\vec{A}[n]}_2 \cdot 
  \card{\vec{b}}_2 ~\leq~ m^{\frac{n+1}{2}} \cdot
  \card{\vec{A}}_{\max}^n \cdot \card{\vec{b}}_\infty\] and the
  conclusion follows, because $(n+1)\cdot m^{\frac{n+1}{2}} \leq
  m^{2n}$, for $n\geq2$. \qed}

%-------------------------------------------------------------------------------
\subsection{Proof of Lemma \ref{lem:omega}}
%-------------------------------------------------------------------------------

\begin{proof}
  ``$\Rightarrow$'' If \((\astate_0,\vec{v}_0) \arrow{\tau_0}{}
  (\astate_1,\vec{v}_1) \arrow{\tau_1}{} \ldots\) is an infinite
  iteration of $c$, then trivially $\vec{v}_i \models \guard{\tau_i}$,
  for all $i \in [0,(\alpha+\beta+1)k-1]$. Moreover, for all $i \in
  [0,k-1]$ we have
  $\vec{C}_i\cdot(\vec{v}_i+p\cdot\vec{w}_i)\leq\vec{d}_i$, for all
  $p\geq0$, which can only hold if $\vec{C}_i\cdot\vec{w}_i \leq 0$.
  ``$\Leftarrow$'' It is sufficient to show that $\vec{v}_i \models
  \guard{\tau_i}$, for all $i \geq (\alpha+\beta)k$. The hypothesis
  ensures then that $\vec{v}_i \models \guard{\tau_i}$ for all $i \in
  \nat$, thus \((\astate_0,\vec{v}_0) \arrow{\tau_0}{} \ldots
  (\astate_1,\vec{v}_1) \arrow{\tau_1}{}\) is an execution. If $i \geq
  (\alpha+\beta+1)k,$, there exist $p\geq1$, $r \in [0,\beta-1]$ and
  $q \in [0,k-1]$ such that $i=(\alpha+p\beta+r)k+q$. By Lemma
  \ref{lem:cycle-translation}, there exist a vector $\vec{w}_q$ such
  that $\vec{v}_i = \vec{v}_{(\alpha+r)k+q} + p\cdot\vec{w}_q$. Then
  $0 \leq (\alpha+r)k+q < (\alpha+\beta+1)k$, thus
  $\vec{v}_{(\alpha+r)k+q} \models \guard{\tau_{(\alpha+r)k+q}}$, by
  the hypothesis and since $\tau_{(\alpha+r)k+q} =\tau_i=\delta_q$ we
  have $\vec{v}_{(\alpha+r)k+q} \models \guard{\delta_q}$. Since $c$
  is infinitely iterable, we know that for each guard
  $\guard{\anedge_q}\equiv\vec{C}_q\cdot\vec{x} \leq \vec{d}_q$, we
  have $\vec{C}_q \cdot \vec{w}_q \leq \vec{0}$, this allows us to
  deduce that $\vec{v}_{(\alpha+r)k+q} + p\cdot\vec{w}_q \models
  \guard{\delta_q}$, for all $p\geq0$. Since $\vec{v}_i =
  \vec{v}_{(\alpha+r)k+q} + p \cdot \vec{w}_q$, we obtain that
  $\vec{v}_i \models \guard{\tau_i}$.\qed
\end{proof}

%-------------------------------------------------------------------------------
\subsection{Proof of Theorem \ref{thm:bounded-iteration}}
%-------------------------------------------------------------------------------

\begin{proof}
We asume that $\wordof{P,\vec{m}}=u_1 c_1^{\ell_1} u_2
c_2^{\ell_2}\ldots u_N c_N^\omega$ where  $u_i \in \Edges^\ast$ and
$c_i$ are the cycles in $P$ for all $i \in [1,N]$. Let $\mathcal{M}_i = \set{\vec{A}_{c_i}^0, \linebreak[0]\ldots,
  \vec{A}_{c_i}^{\alpha_i}, \linebreak[0]\ldots,\linebreak[0]\vec{A}_{c_i}^{\alpha_i+\beta_i}}$
the (finite) monoid of the powers of the update matrix of $c_i$ for each
$i \in [1,N]$,  where
$\alpha_i$ and $\beta_i$ are such that
$\vec{A}_{c_i}^{\alpha_i}=\vec{A}_{c_i}^{\alpha_i+\beta_i}$. We denote
$\alpha=\max_{i=1}^N(\alpha_i+\beta_i)$. By Lemma
\ref{lem:finite-monoid-conditions}, we have $ \alpha \leq 2^{n^3}$.  Moreover, let
$C$ be the maximal length among the cycles $c_1, \ldots, c_N$. Note
that $C \leq \card{\Delta}$ (remember that $\Delta$ is the set of
transition rules). By definition of path schema, we also
have that $N \leq \card{\Edges}$ and $\len{u_1u_2\ldots u_N} \leq \card{\Edges}$ (in a path schema
each element are in fact pairwise distinct).

The first step of the proof consists in rewriting
$\wordof{P,\vec{m}}$ as follows.  Each non-terminal cycle $c_i$ of $P$ that is iterated at
most $\alpha_i+2\beta_i$ times, i.e.\ $\ell_i \leq
\alpha_i+2\beta_i$, is replaced by the finite path $c_i^{\ell_i}$,
obtained by the $\ell_i$-times unfolding of $c_i$. By doing this we
can rewrite $\wordof{P,\vec{m}}$ as $v_1d^{\ell'_1}_1  v_2
d_2^{\ell'_2}\ldots v_Md_M^\omega$, where $v_i \in \Edges^\ast$ for all $i
\in [1,M]$ and $\set{d_1,\ldots,d_M}
\subseteq \set{c_1,\ldots,c_N}$ is the set of cycles in the new path
schema and if we denote by $\gamma_i \in [1,N]$ the unique
number such that $d_i=c_{\gamma_i}$, we have $\ell'_{\gamma_i} >
\alpha_{\gamma_i}+2\beta_{\gamma_i}$.  Since each $v_i$ is obtained from
several rules $u_j$ and unfoldings of simple cycles $c_k$ at most
$\alpha_k + 2\beta_k \leq 2\alpha$ times, this allows us to deduce
that $M \leq \card{\Edges}$ and  $\len{v_j} \leq
2 \alpha .C . \card{\Edges} + \card{\Edges} \leq 3\alpha
.C.\card{\Edges} \leq 3 \alpha \card{\Edges}^2$.

For all $i \in [1,M-1]$, we assume that $\ell'_i=\alpha_{\gamma_i} +
p_i \beta_{\gamma_i}+ r_i$ with $r_i \in [0,\beta_{\gamma_i}-1]$. Also, let $f_{v_i} =
(\vec{A}_{v_i}, \vec{b}_{v_i})$ and $f_{d_i} = (\vec{A}_{d_i},
\vec{b}_{d_i})$ be the affine functions defining the updates of the
finite paths $v_i$ and $d_i$, respectively. In the following, we shall
build a linear system with variables $[\vec{y}; \vec{z}]^\top$, where
$\vec{y}=[\vec{y}_1,\ldots,\vec{y}_n]$ and $\vec{z}=[y_1,\ldots,y_{M-1}]$
  such that $[\vec{v}_0[1],\ldots,\vec{v}_0[n],p_1,\ldots,p_{M-1}]$ is
  a solution of such a system and any solution of the system defines a
  valid run starting at $(q_0,\vec{v}_0)$. 

We now define the terms $\vec{t}_0, \ldots, \vec{t}_{2M+1}$, as
linear combinations of $\vec{y}$ and $\vec{z}$, where $\vec{t}_i =
[t_{i,1}, \ldots, t_{i,n}]$. These terms define the values of the
counters after executing $v_1d_1^{\alpha_{\gamma_1} +
z_1 \beta_{\gamma_1}+ r_1} \ldots v_i$ if
$i$ is odd, and $v_1d_1^{\alpha_{\gamma_1} +
z_1 \beta_{\gamma_1}+ r_1} \ldots
v_id_i^{\alpha_{\gamma_i} +
z_i\beta_{\gamma_i}+ r_i}$ if $i$ is even, respectively. Let
$\vec{t}_0=\vec{y}$ and suppose that we have a definition of
$\vec{t}_{2j}$, for some $j \in [0,M-1]$. Then we define:
\begin{compactitem}
\item $\vec{t}^\top_{2j+1} = f_{v_j}(\vec{t}_{2j}) =
  \vec{A}_{v_j}\cdot\vec{t}^\top_{2j} + \vec{b}_{v_j}$.
\item $\vec{t}^\top_{2j+2}$ is the effect of
  $\alpha_{\gamma_j}+z_j\beta_{\gamma_j}+r_j$ iterations of the simple
  cycle $d_j = c_{\gamma_j}$, starting from the values tracked by
  $\vec{t}_{2j+1}$. By Lemma \ref{lem:cycle-translation}, we obtain
  $\vec{t}^\top_{2j+2} = \vec{A}_{d_j}^{\alpha_{\gamma_j}+r_j} \cdot
  \vec{t}^\top_{2j+1} + \vec{b}_{d_j}^{\alpha_{\gamma_j}+r_j} + y_j\cdot
  \vec{A}_{d_j}^{\alpha_{\gamma_j}} \cdot
  f_{d_j}^{\beta_{\gamma_j}}(\vec{0})$.
\end{compactitem}
This allows us for each $j \in [0,M-1]$ to build two linear systems,
namely $\vec{M}_{v_j}\cdot\vec{t}^\top_{2j} \leq \vec{n}_{v_j}$ and
$\vec{M}_{d_j,r_{j}}\cdot[\vec{t}_{2j+1};z_j]^\top \leq
\vec{n}_{d_j,r_j}$ that capture the fact that all guards have been
satisfied during the execution $v_jd_j^{\alpha_{\gamma_j}+z_j\beta_{\gamma_j}+r_j}$, starting
with $\vec{t}_{2j}$. The last execution $v_M$ is dealt similarly by
a system $\vec{M}_{v_M}\cdot\vec{t}^\top_{2M} \leq \vec{n}_{v_m}$ and
for the last loop we consider the system
$\vec{M}_{d_M,\omega} \cdot \vec{t}_{2M+1} \leq \vec{n}_{d_M,\omega}$,
encoding the fact that $d_M$ can be iterated infinitely many
times. The definitions of the pairs
$(\vec{M}_{d_j,r_j},\vec{n}_{d_j,r_j})$ and $(\vec{M}_{d_M,\omega},
\vec{n}_{d_M,\omega})$ have been describe previously. The definitions of
$(\vec{M}_{v_j}, \vec{n}_{v_j})$ can be produced similarly. 

If we take
the union of these systems we obtain a final system of the shape 
$\vec{M}\cdot[\vec{y};\vec{z}]^\top \leq \vec{n}$ that accepts
$[\vec{v}_0[1],\ldots,\vec{v}_0[n],p_1,\ldots,p_{M-1}]$ as a
solution. And furthermore thanks to the Lemmas \ref{lem:convexity} and
\ref{lem:omega}, we know that if 
$[\vec{v}_0[1],\ldots,\vec{v}_0[n],p'_1,\ldots,\linebreak[0]p'_{M-1}]$ is a
solution of this system, then there exists a run $\rho'$ starting at
$(q_0,\vec{v}_0)$ such that $\wordof{\rho'}=v_1d^{\ell''_1}_1  v_2
d_2^{\ell''_2}\ldots v_Md_M^\omega$ with $\ell''_i=\alpha_{\gamma_i} +
p'_i \beta_{\gamma_i}+ r_i$.

It remains to find bounds for $\card{\vec{M}}_\infty$ and
$\card{\vec{n}}_\infty$. Let $\mu$ be the maximal
value among all $\card{\vec{A}}_\infty$ and $\card{\vec{b}}_\infty$,
where $(\vec{A},\vec{b})$ is any affine function in $S$.  First we
have for each $j \in [1,M]$:

\[\begin{array}{ccc}
\card{\vec{A}_{d_j}}_\infty \leq \mu^C & \hspace*{1cm} & \card{\vec{b}_{d_j}}_\infty \leq C\cdot\mu^C
\end{array}\]

By Lemma \ref{lem:finite-monoid-powers-bound}, we have 
$\card{\vec{A}_{c_i}^{k}}_\infty \leq n \cdot
\card{\vec{A}_{c_i}^{k}}_{\max} \leq n \cdot (n \cdot
\card{\vec{A}_{c_i}}_{\max})^{2n^2} \leq n \cdot n^{2n^2} \cdot
\mu^{2Cn^2} \leq (n\mu)^{2Cn^2+1}$, for each $i \in [1,N]$ and for all $k>0$. Since
$\len{v_j} \leq 3 \alpha \card{\Edges}^2$, we hence
obtain:
\[\begin{array}{ccc}
\card{\vec{A}_{v_j}}_\infty \leq \mu^{3 \alpha \card{\Edges}^2} \cdot
    (n\mu)^{(2C n^2 +1).3 \alpha \card{\Edges}^2} & \hspace*{1cm} &
                                                                    \card{\vec{b}_{v_j}}_\infty
                                                                    \leq
                                                                    3
                                                                    \alpha
                                                                    \card{\Edges}^2.(
                                                                    \mu+
                                                                    (n\mu)^{2Cn^2+1})
\end{array}\]
We have also
\[
\card{\vec{A}_{d_i}^{\alpha_{\gamma_i}+r_i}}_\infty  \leq 
                                                             (n\mu)^{2Cn^2+1}
                                                             \\
\]
and since $\alpha_{\gamma_i}+ \beta_{\gamma_i} \leq \alpha$, for $r_i < \beta_{\gamma_i}$, we
deduce:
\[\begin{array}{rcccl}
\card{\vec{b}_{d_i}^{\alpha_{\gamma_i}+r_i}}_\infty & \leq & (\alpha_{\gamma_i}+r_i)  C\cdot\mu^C
& \leq & \alpha\cdot C\cdot\mu^C \\
\end{array}
\]
Last, 
\[\begin{array}{rcccl}
\card{\vec{A}_{d_i}^{\alpha_{\gamma_i}}\cdot
    f_{d_i}^{\beta_{\gamma_i}}(0)}_\infty & \leq &
                                                   \card{\vec{A}_{d_i}^{\alpha_{\gamma_i}}}_\infty
                                                   \cdot
                                                   \beta_{\gamma_i}.
                                                   C.\mu^C.(n\mu)^{2Cn^2+1}
& \leq & \alpha C.\mu^C.(n\mu)^{4Cn^2+2}
\end{array}\]

Now observe that, for all $i \in
[0,2M+1]$, we have $\vec{t}_i = \vec{U}_i \cdot \vec[\vec{y};
  \vec{z}]^\top + \vec{v}_i$, for some $\vec{U}_i \in \zed^{n \times
  (n+M-1)}$ and $\vec{v}_i \in \zed^n$. 
By the definition of the different term $\vec{t}_i$, we deduce  for all $j \in  [0,M-1]$,
\[\begin{array}{rcl}
\card{\vec{U}_{2j+1}}_\infty & \leq & \card{\vec{A}_{v_j}}_\infty \cdot \card{\vec{U}_{2j}}_\infty \\
\card{\vec{v}_{2j+1}}_\infty & \leq & \card{\vec{A}_{v_j}}_\infty \cdot \card{\vec{v}_{2j}}_\infty + \card{\vec{b}_{v_j}}_\infty  \\
\card{\vec{U}_{2j+2}}_\infty & \leq & \card{\vec{A}_{d_j}^{\alpha_{\gamma_j}+r_j}}_\infty \cdot \card{\vec{U}_{2j+1}}_\infty + \card{\vec{A}_{d_i}^{\alpha_{\gamma_i}}\cdot f_{d_i}^{\beta_{\gamma_i}}(0)}_\infty \\
\card{\vec{v}_{2j+2}}_\infty & \leq & \card{\vec{A}_{d_i}^{\alpha_{\gamma_i}+r_i}}_\infty \cdot \card{\vec{v}_{2j+1}}_\infty + \card{\vec{b}_{d_i}^{\alpha_{\gamma_i}+r_i}}_\infty \\
\end{array}\]

Since $j \in [0,M]$ and $M \leq N$ and  $ \alpha \leq 2^{n^3}$,  we obtain that
$\card{\vec{U}_i}_\infty$ and $\card{\vec{v}_i}_\infty$ are bounded by
$2^{\polyone{\sizeof{S}}}$, for a suitable polynomial function
$\polyone{x}$. Since the  coefficients of the linear system
$\vec{M}\cdot[\vec{x};\vec{y}]^\top \leq \vec{n}$ are the coefficients
of products between $\vec{C}_j$ and $\vec{U}_i$, respectively
$\vec{C}_j$ and $\vec{v}_i$, for a guard $\vec{C}_i \cdot \vec{x} \leq
\vec{d}_i$ of $S$, we deduce that the coefficient in the linear system
are bounded by
$2^{\polytwo{\sizeof{S}}}$ for a polynomial function
$\polytwo{x}$. Moreover, the number of rows in $\vec{M}$ is equals to
$\Sigma_{j \in [1,M-1]} \len{v_j} +
(\len{d_j}.(\alpha_{\gamma_j}+\beta_{\gamma_j} + 1 + \beta_{\gamma_j}))+
\len{v_M}  +\len{d_M}.(\alpha_{\gamma_M}+\beta_{\gamma_M} + 1)$ which
is less than $(M+1).\card{\Edges}.2 \alpha \leq
\card{\Edges^2}.2.2^{n^3}$ which yields a $2^{\polythree{\sizeof{S}}}$
bound on the number of rows in the system.

As mentionned before
$[\vec{v}_0[1],\ldots,\vec{v}_0[n],p_1,\ldots,p_{M-1}]$ is an integral
solution of the system $\vec{M}\cdot[\vec{y};\vec{z}]^\top \leq
\vec{n}$, by  application of Theorem
\ref{thm:ineq-sol-bound} we obtain that $\vec{M} \cdot [\vec{x};
  \vec{y}]^\top \leq \vec{n}$ has also a solution
  $[\vec{v}_0[1],\ldots,\vec{v}_0[n],p'_1,\ldots,p'_{M-1}]$ such  $0 \leq p'_i \leq
2^{\poly{\sizeof{S}+\sizeof{\vec{v}_0}}}$ for all $i \in [0.M-1]$. But
as said before then there exists a run $\rho'$ starting at
$(q_0,\vec{v}_0)$ such that $\wordof{\rho'}=v_1d^{\ell''_1}_1  v_2
d_2^{\ell''_2}\ldots v_Md_M^\omega$ with $\ell''_i=\alpha_{\gamma_i} +
p'_i \beta_{\gamma_i}+ r_i$. By folding back this $\omega$-word to
match it with the path schema $P$, we obtain a vector $\vec{m}'$ such that
$\tuple{P,\vec{m'}} \in \ips{\arun'}$ and $\card{\vec{m}'}_\infty \leq
2^{\poly{\sizeof{S}+\sizeof{\vec{v}_0}}}$. \qed

\end{proof}

%-------------------------------------------------------------------------------
\subsection{Proof of Lemma \ref{lem:co-np-oracle}}
%-------------------------------------------------------------------------------

\begin{proof}
Let $\tuple{P,\vec{m}}$ be an iterated path schema of a counter system
  with the finite monoid property $S$ and
 $\gamma_0=(\astate_0,\vec{v}_0)$ an initial configuration.
We suppose that the cycles of $P$ are $c_1,\ldots,c_N$ and 
for each $i \in [1,N]$, let $\set{\vec{A}_{c_i}^0, \ldots,
  \vec{A}_{c_i}^{\alpha_i}, \ldots, \vec{A}_{c_i}^{\alpha_i+\beta_i}}$
be the finite monoid of the powers of the update matrix of $c_i$,
where $\alpha_i$ and $\beta_i$ are such that
$\vec{A}_{c_i}^{\alpha_i}=\vec{A}_{c_i}^{\alpha_i+\beta_i}$. 

First note that since in pseudo-run we do not test the values of the
counters, there exists exactly one pseudo-run $\arun :
(\astate_0,\vec{v}_0) \stackrel{\scriptscriptstyle \tau_0}{\leadsto}
(\astate_1,\vec{v}_1) \stackrel{\scriptscriptstyle \tau_1}{\leadsto}
\ldots $ such that $\wordof{\arun} = \wordof{P,\vec{m}}$. We
deduce that there is no run $\rho'$ starting at  $\gamma_0$
  such that $\tuple{P,\vec{m}} \in \ips{\arun'}$ iff one guard is not
  satisfied in the pseudo-run $\arun$. We will hence look for a
  position $p$ in the $\arun$ such that $\vec{v}_p \not\models
  \guard{\tau_p}$.

First we check that the last loop $c_N$ is infinitely iterable. We recall that
$c_N$ is infinitely iterable iff $\bigwedge_{i \in [0,
    \len{c_N}-1]}\vec{C}_N \cdot \vec{w}_i \leq 0$, where $\vec{w}_0 =
\vec{A}_{c_N}^{\alpha_N} \cdot f_{c_N}^{\beta_N}(\vec{0})$, and
$\vec{w}_{i+1} = \vec{A}_{c_{N,i}} \cdot \vec{w}_i$. Using the bounds
from the proof of Theorem \ref{thm:bounded-iteration}, and the fact
that $\alpha_N + \beta_N\leq 2^{n^3}$ (Lemma
\ref{lem:finite-monoid-bound}), we infer that the sizes of $\vec{w}_i$
are polynomially bounded by $\sizeof{P}$, thus the test $\bigwedge_{i
  \in [0, \len{c_N}-1]}\vec{C}_N \cdot \vec{w}_i \leq 0$ takes
polynomial time. If $c_N$ is not infinitely iterable we know that there is $p$ such that $\vec{v}_p \not \models \guard{\tau_p}$.

On the other hand if $c_N$ is infintely iterable, we know (thanks to
Lemma \ref{lem:omega}), that if  there is $p$ such that $\vec{v}_p \not
\models \guard{\tau_p}$, then $p \leq |P[1]^{\vec{m}[1]}P[2]^{\vec{m}[2]} ~\cdots~
P[\lengthof{P}-1]^{\vec{m}[\lengthof{P}-1]}P[\lengthof{P}]^{\alpha_N+\beta_N+1}|$. Hence
if $c_N$ is infintely iterable, we guess $p \leq |P[1]^{\vec{m}[1]}P[2]^{\vec{m}[2]} ~\cdots~
P[\lengthof{P}-1]^{\vec{m}[\lengthof{P}-1]}P[\lengthof{P}]^{\alpha_N+\beta_N+1}|$,
and we know that $\vec{v}_p$ can be computed in time
polynomially bounded by \(\sizeof{P} + \sizeof{\vec{m}} +
\sizeof{\vec{v}_0}\), using exponentiation by squaring. Consequently,
$\sizeof{\vec{v}_p}$ is polynomially bounded by \(\sizeof{P} +
\sizeof{\vec{m}} + \sizeof{\vec{v}_0}\), and checking whether$\vec{v}_p
\models \guard{\tau_p}$ takes polynomial time in the size of
the input. 

This gives an \textsc{NP} algorithm that decides whether the pseudo-run
$\arun$ is not a run. \qed

\end{proof}

%-------------------------------------------------------------------------------
\subsection{Proof of Theorem \ref{cor:reach-sigma-2-p}}
%-------------------------------------------------------------------------------

\begin{proof}
By Lemma \ref{lemma:reach-hardness}, $\Reach(\FlatMonoidAffineSystems)$
is $\Sigma_2^P$-hard. To show inclusion in $\Sigma_2^p$, we prove the
existence of a $\textsc{NP}^\textsc{NP}$ algorithm deciding any
instance of $\Reach(\FlatMonoidAffineSystems)$. Let $S =
\tuple{\States,\Counters_n,\Edges,\lab}$ be a flat affine counter
system, with the finite monoid property, and \((\astate_0,\vec{v}_0)\)
be an initial configuration of $S$. If $S$ has a run $\arun$, starting
in \((\astate_0,\vec{v}_0)\), then there exists an iterated path
schema $\tuple{P,\vec{m}} \in \ips{\arun}$ such that $\sizeof{P}$ is
polynomially bounded by $\sizeof{S}$ (Lemma
\ref{lem:iterated-path-schema}). By Theorem
\ref{thm:bounded-iteration}, there exists also a run $\arun'$,
starting in \((\astate_0,\vec{v}_0)\), and an iterated path schema
$\tuple{P,\vec{m}'} \in \ips{\arun'}$, such that
$\card{\vec{m}'}_\infty \leq
2^{\poly{\sizeof{S}+\sizeof{\vec{v}_0}}}$, for a suitable polynomial
function $\poly{x}$. Then we consider a nondeterministic procedure
that guesses the entries of $\vec{m}'$ in time
$\poly{\sizeof{S}+\sizeof{\vec{v}_0}}$, and an oracle that checks
whether the guessed values correspond to a real execution of $S$. By
Lemma \ref{lem:co-np-oracle}, this test can be implemented by an
algorithm that is \textsc{NP} in
$\sizeof{P}+\sizeof{\vec{m}'}+\sizeof{\vec{v}_0}$, and consequently,
in $\textsc{NP}$ in $\sizeof{S}+\sizeof{\vec{v}_0}$. We have
obtained a $\textsc{NP}^{\textsc{NP}}$ algorithm that decides any instance
of $\Reach(\FlatMonoidAffineSystems)$. Therefore
$\Reach(\FlatMonoidAffineSystems)$ belongs to $\Sigma_2^P$. \qed
\end{proof}

%-------------------------------------------------------------------------------
\subsection{Proof of Theorem \ref{thm:model-checking-pltl}}
%-------------------------------------------------------------------------------

\begin{proof}
We consider a flat counter system with the finite monoid property $S =
\tuple{\States,\Counters_n,\Edges,\lab}$, an initial configuration
$\aconf_0=(\astate_0,\vec{v}_0)$ and a $\PLTL$ formula $\phi$. 

Assume there is a run $\rho$
starting at $\aconf_0$ such that $\arun \modelspltl \phi$. Then from
Lemma \ref{lem:iterated-path-schema}  there exists an iterated path
schema $\tuple{P,\vec{m}} \in \ips{\arun}$ such that $\sizeof{P}$ is
polynomially bounded by $\sizeof{S}$. From Lemma
\ref{lem:bounded-iteration-pltl}, we know that there exists a run \(\arun'\),
  starting in $\aconf_0$, and an iterated path schema
  $\tuple{P,\vec{m}'} \in \ips{\arun'}$, such that
  $\card{\vec{m}'}_\infty \leq
  2^{\poly{\sizeof{S}+\sizeof{\vec{v}_0}+td(\phi)}}$ for a polynomial function
  $\poly{x}$ and $\xi_{2td(\phi)+5}(\vec{m})=\xi_{2td(\phi)+5}(\vec{m'})$, By definition,
  note that  $\lab(P,\xi_{2
    td(\phi)+5}(\vec{m}))=\lab(P,\xi_{2
    td(\phi)+5}(\vec{m'}))$. Since $\arun \modelspltl \phi$, we deduce
  that $\lab(P,\vec{m}),0 \modelspltl \phi$ and as consequence of
  Lemma \ref{lemma:pltl-path-schema}.1, we deduce that
    $\lab(P,\xi_{2
    td(\phi)+5}(\vec{m})),0 \modelspltl \phi$. This allows us to deduce
    that there exists a path schema $\tuple{P,\vec{m}'}$ such that
    $\card{\vec{m}'}_\infty \leq
    2^{\poly{\sizeof{S}+\sizeof{\vec{v}_0}+td(\phi)}}$ and $\lab(P,\xi_{2
    td(\phi)+5}(\vec{m'})),0 \modelspltl \phi$ and there exists a run $\rho'$
  starting at $\aconf_0$such that $\tuple{P,\vec{m}'}
  \in \ips{\arun'}$. This guarantees that our algorithm is sound.

The correctness of our algorithm is ensures by the fact that if we
find an iterated path schema $\tuple{P,\vec{m}}$ such that $\lab(P,\xi_{2
    td(\phi)+5}(\vec{m})),0 \modelspltl \phi$, then we know from Lemma
  \ref{lemma:pltl-path-schema}.1 that   $\lab(P,\vec{m}),0
  \modelspltl \phi$ and consequently any run $\rho$ starting in
  $\aconf_0$ such that $\tuple{P,\vec{m}}
  \in \ips{\arun}$ verifies that $\arun \modelspltl \phi$.

Finally the fact that our algorithm is $\Sigma_2^P$ is because we 
guess  an iterated path schema $\tuple{P,\vec{m}}$ of polynomial size
  in the size of $S$. $\vec{v}_0$ and $\phi$. Then checking that $\lab(P,\xi_{2
    td(\phi)+5}(\vec{m})),0 \modelspltl \phi$ can be done in
  polynomial time in the size of $P$ and $\phi$ thanks to Lemma
  \ref{lemma:pltl-path-schema}.2 and verifying whether there exists or
  not a run $\rho$ starting at $\aconf_0$, such
that $\tuple{P,\vec{m}} \in \ips{\rho}$ can be done thanks to the
\textsc{NP} algorithm given in Lemma \ref{lem:co-np-oracle} 
\end{proof}

%-------------------------------------------------------------------------------
\subsection{Proof of Theorem \ref{thm:model-checking-fo}}
%-------------------------------------------------------------------------------

First we need to state the pendant of Lemma
\ref{lemma:pltl-path-schema} but for $\FO$ formulae. Again this result
can be proved by adapting the proof of Theorem
\ref{thm:bounded-iteration} unfolding the loop $2^{qh(\phi)+2}$ for a
considered $\FO$ formula $\phi$. Note however that even if we unfold
in this case
some loops an exponential number of times, it was already the case for
some loops that we had to unfold a number of times equal to the size
of finite monoid of the counter system.

\begin{lemma}\label{lem:bounded-iteration-fo}
  For any run \(\arun\) of $S$, starting in
  \((\astate_0,\vec{v}_0)\), and any iterated path schema
  $\tuple{P,\vec{m}} \in \ips{\arun}$ and any $\FO$
formula $\phi$, there exists a run \(\arun'\),
  starting in \((\astate_0,\vec{v}_0)\), and an iterated path schema
  $\tuple{P,\vec{m}'} \in \ips{\arun'}$, such that
  $\card{\vec{m}'}_\infty \leq
  2^{\poly{\sizeof{S}+\sizeof{\vec{v}_0}+qh(\phi)}}$ for a polynomial function
  $\poly{x}$ and $\xi_{2^{qh\phi)+2}}(\vec{m})=\xi_{2^{qh(\phi)+2}}(\vec{m'})$.
\end{lemma}

As a consequence of Lemma \ref{lem:fo-path-schema}, the new run $\rho'$
obtained in the previous lemma is such that $\rho \modelsfo \phi$
iff $\rho' \modelsfo \phi$ for the considered $\FO$ formula $\phi$.

Then the proof follows essentially the same step as the proof of
Theorem \ref{thm:model-checking-fo}. The fact that our algorithm is in
\textsc{PSPACE} comes from the fact that at some point we will need to
check that $\lab(P,\xi_{2^{qh(\phi)+2}}(\vec{m})) \modelsfo \phi$ for
an iterated path schema $\tuple{P,\vec{m}}$ and this can only be done
in polynomially space. In reality our algorithm is in
\textsc{NPSPACE} but by Savitch's theorem, we have that
\textsc{NPSPACE} is equal to \textsc{PSPACE}.

%%%%%%%%%%%%%%%%%%%%%%%%%%%%%%%%%%%%%%%%%%%%%%%%%%%%%%%%%%%%%%%%%%%%%%%%%%%%%%%%
\section{Additional Technical Results}\label{app:algebra}
%%%%%%%%%%%%%%%%%%%%%%%%%%%%%%%%%%%%%%%%%%%%%%%%%%%%%%%%%%%%%%%%%%%%%%%%%%%%%%%%

%-------------------------------------------------------------------------------
\subsection{Bounding the Cardinality and Values of a Finite Monoid}
%-------------------------------------------------------------------------------

Some basic notions of linear algebra are needed in the following. If
$\vec{A} \in \zed^{n \times n}$ is a square matrix, and $\vec{v} \in
\zed^n$ is a column vector, any complex number $\lambda \in \complex$,
such that $\vec{A} \vec{v} = \lambda \vec{v}$, for some vector
$\vec{v} \in \complex^n$, is called an \emph{eigenvalue} of
$\vec{A}$. The vector $\vec{v}$ in this case is called an
\emph{eigenvector} of $\vec{A}$. The eigenvalues of $\vec{A}$ are the
roots of the \emph{characteristic polynomial}
$P_{\vec{A}}(x)=\mbox{det}(\vec{A} - x \idmatrix_n)$. If
$\lambda_1,\ldots,\lambda_m$ are the eigenvalues of $\vec{A}$, then
$\lambda_1^p, \ldots, \lambda_m^p$ are the eigenvalues of $\vec{A}^p$,
for all $p > 0$.

A matrix $\vec{A} \in \zed^{n \times n}$ is said to be
\emph{diagonalizable} iff there exists a non-singular matrix $\vec{U}
\in \complex^{n \times n}$ and a diagonal matrix $\vec{D} \in
\complex^{n \times n}$, with the eigenvalues of $\vec{A}$ occurring on
the main diagonal, such that $\vec{A} = \vec{U} \cdot \vec{D} \cdot
\vec{U}^{-1}$. A \emph{Jordan block} $\vec{B}$ of $\vec{A}$ is a
square matrix of dimension $k \leq n$ with an eigenvalue of $\vec{A}$
on the main diagonal, the constant $1$ above the main diagonal, and
zero everywhere else. Every matrix $\vec{A}$ can be written as
$\vec{A}= \vec{V} \cdot \vec{J} \cdot \vec{V}^{-1}$, where $\vec{V}
\in \complex^{n \times n}$ is a non-singular matrix consisting of the
eigenvectors of $\vec{A}$ and $\vec{J} \in \complex^{n \times n}$ has
Jordan blocks on its main diagonal and zero everywhere else. A matrix
is diagonalizable iff each of its Jordan blocks is of dimension one.

A complex number $r \in \complex$ is said to be a \emph{root of the
  unity} if $r^d = 1$ for some integer $d > 0$. The \emph{cyclotomic
  polynomial} $F_d(x)$ is the product of all monomials $x - \omega$,
where $\omega^d = 1$, and $\omega^e \neq 1$, for all $0 < e < d$. A
polynomial has only roots which are roots of unity iff it is a product
of cyclotomic polynomials.

\begin{lemma}\label{lem:finite-monoid-conditions}\cite{BoigelotPhD,FinkelLeroux02}
  Given a matrix $\vec{A} \in \zed^{n \times n}$, the monoid
  $\mathcal{A}=\set{\vec{A}^0,\vec{A}^1,\vec{A}^2,\ldots}$ of powers
  of $\vec{A}$ is finite iff there exists $p>0$ such that the
  following hold: \begin{compactenum}
  \item\label{it1:finite-monoid-conditions} the eigenvalues of $\vec{A}^p$
    belong to the set $\set{0,1}$, and
  \item\label{it2:finite-monoid-conditions} $\vec{A}^p$ is diagonalizable.
  \end{compactenum}
\end{lemma}
\proof{Initially proved as \cite[Theorems 8.42 and
    8.44]{BoigelotPhD}. A simpler proof can be found in \cite[Lemma
    1]{FinkelLeroux02}. \qed}

The following lemma shows that, given a matrix $\vec{A} \in \zed^{n \times
  n}$, if the monoid of the powers of $\vec{A}$ is finite, then its
cardinality is bounded by a simple exponential in the dimension $n$ of
the matrix. Note that it does not depend on the entries of $\vec{A}$,
because the finite monoid condition imposes rather strict restrictions
on the values occuring in $\vec{A}$ (Lemma
\ref{lem:finite-monoid-conditions}
(\ref{it1:finite-monoid-conditions})).

\begin{lemma}\label{lem:finite-monoid-bound}
  Given a matrix $\vec{A} \in \zed^{n \times n}$, for some $n\geq2$,
  the monoid $\amonoid= \set{\vec{A}^0,\vec{A}^1,\vec{A}^2,\ldots}$ of
  powers of $\vec{A}$ is finite if and only if $\card{\amonoid} \leq
  2^{n^3}$.
\end{lemma}
\begin{proof}
  The direction from right to left is trivial. For the other
  direction, assume that $\amonoid$ is finite. By Lemma
  \ref{lem:finite-monoid-conditions}, there exists $p>0$ such that
  $\vec{A}^p = \vec{U} \cdot \vec{D} \cdot \vec{U}^{-1}$. We have
  $\vec{A}^p=\vec{A}^{2p}$, thus $\card{\amonoid} \leq 2p$. For a
  bound on $p$, observe that, since for any eigenvalue $\lambda$ of
  $\vec{A}$, we have that $\lambda^p$ is an eigenvalue of $\vec{A}^p$,
  then $\lambda$ must be either zero or a root of the unity. Then the
  characteristic polynomial $P_{\vec{A}}$ of $\vec{A}$ is a product of
  $x^m$, for some $m < n$, and several cyclotomic polynomials, call
  them $\Phi_{i_1},\ldots,\Phi_{i_q}$. Let $\ell =
  \lcm(i_1,\ldots,i_q)$. First, we have that $\lambda^{k\ell}=1$ for
  each $k>0$ and each non-zero eigenvalue $\lambda$ of
  $\vec{A}$. Suppose that $\vec{A}^\ell$ is not diagonalizable, then
  it has one or more Jordan blocks $\vec{B}_1,\ldots,\vec{B}_h$ of
  dimension two or more, with eigenvalues
  $\lambda_1,\ldots,\lambda_h$, on the main diagonal, respectively. If
  $\lambda_j=1$ for some $j \in [1,h]$, then the set of powers of
  $B_j$ is not finite, thus the monoid of powers of $\vec{A}$ cannot
  be finite either, which contradicts the hypothesis of the lemma.
  Hence, it must be the case that $\lambda_1=\ldots=\lambda_h=0$, thus
  $\vec{B}_1^2 = \ldots = \vec{B}_h^2$ have zero entries everywhere,
  and consequently, $\vec{A}^{2\ell}$ is the zero matrix. In any case
  --- either $\vec{A}^\ell$ is diagonalizable or not --- we find that
  $2\ell$ meets the conditions (\ref{it1:finite-monoid-conditions})
  and (\ref{it2:finite-monoid-conditions}) of Lemma
  \ref{lem:finite-monoid-conditions}, thus $p \leq 2\ell$. To find a
  bound on $\ell$, we use the fact that, for any $k>0$ and $d \geq 0$,
  such that $k > 210 \left(\frac{d}{48}\right)^{\log_{10}11}$, the
  degree of the cyclotomic polynomial $\Phi_k(x)$ is higher than $d$
  \cite[Theorem 8.46]{BoigelotPhD}. Since $i_1,\ldots,i_q$ are the
  indices of cyclotomic polynomials $\Phi_{i_1}, \ldots, \Phi_{i_q}$
  dividing $P_{\vec{A}}$, their degrees are less than or equal to $n$,
  thus $i_1, \ldots, i_q \leq 210
  \left(\frac{n}{48}\right)^{\log_{10}11} \leq n^2$. Then
  $\ell=\lcm(i_1,\ldots,i_q) \leq n^{2n^2} \leq 2^{n^3}$. \qed
\end{proof}

The next lemma bounds the size of the binary representation of any
power $\vec{A}^k$ of a finite monoid matrix $\vec{A}$ by a polynomial
function of $\sizeof{\vec{A}}$. 

\begin{lemma}\label{lem:finite-monoid-powers-bound}
  Given a matrix $\vec{A} \in \zed^{n \times n}$, $n\geq2$, such that
  the monoid $\amonoid= \set{\vec{A}^0,\vec{A}^1,\vec{A}^2,\ldots}$ is
  finite, for any $k>0$, we have $\card{\vec{A}^k}_{\max} \leq
  (n\cdot\card{\vec{A}_{\max}})^{2n^2}$.
%%  and $\sizeof{\vec{A}^k} \leq 3n^5 \cdot \sizeof{\vec{A}}$.
\end{lemma}
\begin{proof}
Let $\vec{A} = \vec{D} \cdot \vec{J} \cdot \vec{D}^{-1}$ be a Jordan
decomposition of $A$, where $\vec{D} \in \complex^{n \times n}$ is a
non-singular matrix consisting of eigenvectors of $\vec{A}$ and
$\vec{J}$ consists of Jordan blocks $\vec{B}_1, \ldots, \vec{B}_h$,
with eigenvalues $\lambda_1, \ldots, \lambda_h \in \complex$,
respectively. By the argument used in the proof of Lemma
\ref{lem:finite-monoid-bound}, $\lambda_1, \ldots, \lambda_h$ are
either $0$ or roots of unity. Since $\vec{A}^k = \vec{D} \cdot
\vec{J}^k \cdot \vec{D}^{-1}$, we have $\card{\vec{A}^k}_\infty \leq
\card{\vec{D}}_\infty \cdot \card{\vec{J}^k}_\infty \cdot
\card{\vec{D}^{-1}}_\infty$. In the following, we compute bounds on
$\card{\vec{D}}_\infty$, $\card{\vec{J}^k}_\infty$ and
$\card{\vec{D}^{-1}}_\infty$, respectively: \begin{compactenum}
\item Since the columns of $\vec{D}$ are non-zero solutions of a
  homogeneous linear equation
  $(\vec{A}-\lambda_i\idmatrix_n)\vec{x}=\vec{0}$, for some $i \in
  [1,h]$, such that $\det(\vec{A}-\lambda_i\idmatrix_n)=0$, we can
  factorize $\vec{x} = [x_1;\ldots;x_\ell;x_{\ell+1};\ldots;x_n]^\top$
  (if necessary, by a permutation of the columns of $\vec{A}$) such
  that $x_{\ell+1} = \ldots = x_n=1$ (these values can be chosen
  arbitrarily) and $x_1,\ldots, x_\ell$ are solutions of a system
  $\vec{A}' [x_1 \ldots x_\ell]^\top = \vec{b}$, where
  $[\vec{A}';\vec{b}]$ is a sub-matrix of $\vec{A}$, with $\vec{A}'$ a
  singular square matrix. Then, the solutions $x_1, \ldots, x_\ell$
  are bounded by the sub-determinants of $\vec{A}$. By Hadamard's inequality, we have:
  \[\begin{array}{rcl}
  \card{\vec{D}}_{\max} & \leq & \card{\vec{A}[1]}_2 \cdot \ldots \cdot \card{\vec{A}[n]}_2 \\
  & \leq & n^{\frac{n}{2}} \card{A}_{\max}^n \text{, because $\card{\vec{A}[i]}_2 \leq \sqrt{n}\card{A}_{\max}$}
  \end{array}\]
\item Since $\vec{D} \cdot \vec{D}^{-1} = \idmatrix_n$, the columns of
  $\vec{D}^{-1}$ are the unique solutions of the systems $\vec{D}
  \cdot \vec{x} = \vec{e}_i$, where $\vec{e}_i$ is the column vector
  with $1$ on the $i$-th row and zero everywhere else. By the same
  argument as before, we obtain: 
  \[\begin{array}{rcl}
  \card{\vec{D}^{-1}}_{\max} & \leq & \card{\vec{D}[1]}_2 \cdot \ldots \cdot \card{\vec{D}[n]}_2 
  ~\leq~ n^{\frac{n}{2}} \cdot \card{\vec{D}}_{\max}^n \\
  & \leq & n^{\frac{n^2}{2} + \frac{n}{2}} \card{\vec{A}}_{\max}^{n^2} 
  \text{, because $\card{\vec{D}}_{\max} \leq n^{\frac{n}{2}} \card{A}_{\max}^n$}
  \end{array}\]
\item For each Jordan block $\vec{B}_i$ of $A$, $i \in [1,h]$, we
  distinguish three cases: \begin{compactenum}[(i)]
  \item the dimension of $\vec{B}_i$ is one and $\lambda_i$ is a root of the unity,
  \item the dimension of $\vec{B}_i$ is more than one and
    $\lambda_i=0$; in this case $\vec{B}_i^k = \vec{0}$, for all
    $k>1$,
  \item the dimension of $\vec{B}_i$ is more than one and $\lambda_i$
    is a root of the unity. In this case the set $\set{\vec{B}_i^k
      \mid k > 0}$ is not finite, thus $\set{\vec{J}^k \mid k > 0}$
    and, consequently $\mathcal{M} = \set{\vec{A}^k \mid k > 0}$ are
    not finite, which is in contradiction with the hypothesis of the
    lemma. 
  \end{compactenum}
  Then for all $k>0$ and $i \in [1,h]$, we have
  $\card{\vec{B}^k_i}_\infty \leq 1$, thus $\card{\vec{J}^k}_\infty
  \leq 1$ and:
\end{compactenum} 
\[\begin{array}{rcl}
\card{\vec{A}^k}_{\max} ~\leq~ \card{\vec{A}^k}_\infty & \leq & 
\card{\vec{D}}_\infty \cdot \card{\vec{J}^k}_\infty \cdot \card{\vec{D}^{-1}}_\infty \\
& \leq & n^2\cdot\card{\vec{D}}_{\max} \cdot \card{\vec{J}^k}_\infty \cdot \card{\vec{D}^{-1}}_{\max} \\
& \leq & n^2\cdot n^{\frac{n^2}{2} + \frac{n}{2} + 1} \cdot \card{\vec{A}}_{\max}^{n^2+n} \\ 
& \leq & (n\cdot\card{\vec{A}}_{\max})^{2n^2}
\end{array}\]
%% And, consequently: 
%% \[\begin{array}{rcl}
%% \sizeof{A^k} & \leq & n^2 \log \card{A^k}_{\max} \\
%% & \leq & n^2(n^2 \log_2n + (n^2+n)\log_2\card{\vec{A}_{\max}}) \\
%% & \leq & 3n^5 \log_2\card{\vec{A}_{\max}} \\
%% & \leq & 3n^5 \cdot \sizeof{\vec{A}}\enspace.
%% \end{array}\]
\qed
\end{proof}

\end{document}